\documentclass[acmtog]{acmart}

\AtBeginDocument{%
  \providecommand\BibTeX{{%
    \normalfont B\kern-0.5em{\scshape i\kern-0.25em b}\kern-0.8em\TeX}}}

\setcopyright{acmlicensed}
\acmJournal{TOG}
\acmYear{2022} \acmVolume{41} \acmNumber{3} \acmArticle{33} \acmMonth{6} \acmDOI{10.1145/3516429}

\usepackage{booktabs} 
\usepackage{subcaption}
\usepackage{multicol}
\usepackage{multirow}

\usepackage[ruled]{algorithm2e} 

\SetAlFnt{\small}
\SetAlCapFnt{\small}
\SetAlCapNameFnt{\small}
\SetAlCapHSkip{0pt}

\usepackage{kotex}
\usepackage{enumitem}
\usepackage{color}
\definecolor{blue}{rgb}{0,0,0.8}
\definecolor{green}{rgb}{0,0.6,0}
\definecolor{red}{rgb}{0.7,0,0}

\usepackage[normalem]{ulem}

\def\mat#1{\mathbf{#1}}
\def\vec#1{\mathbf{#1}}

\newcommand{\vf} {\vec{f}}

\newcommand{\vm} {\vec{m}}

\newcommand{\vr} {\vec{r}}

\newcommand{\vw} {\vec{w}}

\newcommand \mM	{\mat{M}}

\newcommand{\real} {\mathbb{R}}

\begin{document}

\title{Motion Puzzle: Arbitrary Motion Style Transfer by Body Part}

\thanks{This work was supported by National Research Foundation, Korea (NRF-2020R1A2C2011541), and Korea Creative Content Agency, Korea (R2020040211).}

\author{Deok-Kyeong Jang}
\email{shadofex@kaist.ac.kr}
\affiliation{%
  \institution{Korea Advanced Institute of Science and Technology}
  \streetaddress{291 Daehak-ro, Yuseong-gu}
  \city{Daejeon}
  \postcode{34141}
  \country{South Korea}
}

\author{Soomin Park}
\email{sumny@kaist.ac.kr}
\affiliation{%
  \institution{Korea Advanced Institute of Science and Technology}
  \country{South Korea}
}

\author{Sung-Hee Lee}
\email{sunghee.lee@kaist.ac.kr}
\affiliation{%
  \institution{Korea Advanced Institute of Science and Technology}
  \country{South Korea}
}


\begin{abstract}
This paper presents Motion Puzzle, a novel motion style transfer network that advances the state-of-the-art in several important respects. The Motion Puzzle is the first that can control the motion style of individual body parts, allowing for local style editing and significantly increasing the range of stylized motions. Designed to keep the human's kinematic structure, our framework extracts style features from multiple style motions for different body parts and transfers them locally to the target body parts. 
Another major advantage is that it can transfer both global and local traits of motion style by integrating the adaptive instance normalization and attention modules while keeping the skeleton topology. Thus, it can capture styles exhibited by dynamic movements, such as flapping and staggering, significantly better than previous work. 
In addition, our framework allows for arbitrary motion style transfer without datasets with style labeling or motion pairing, making many publicly available motion datasets available for training. 
Our framework can be easily integrated with motion generation frameworks to create many applications, such as real-time motion transfer. 
We demonstrate the advantages of our framework with a number of examples and comparisons with previous work.

\end{abstract}



\begin{teaserfigure}
  \centering
  \includegraphics[width=6.5in]{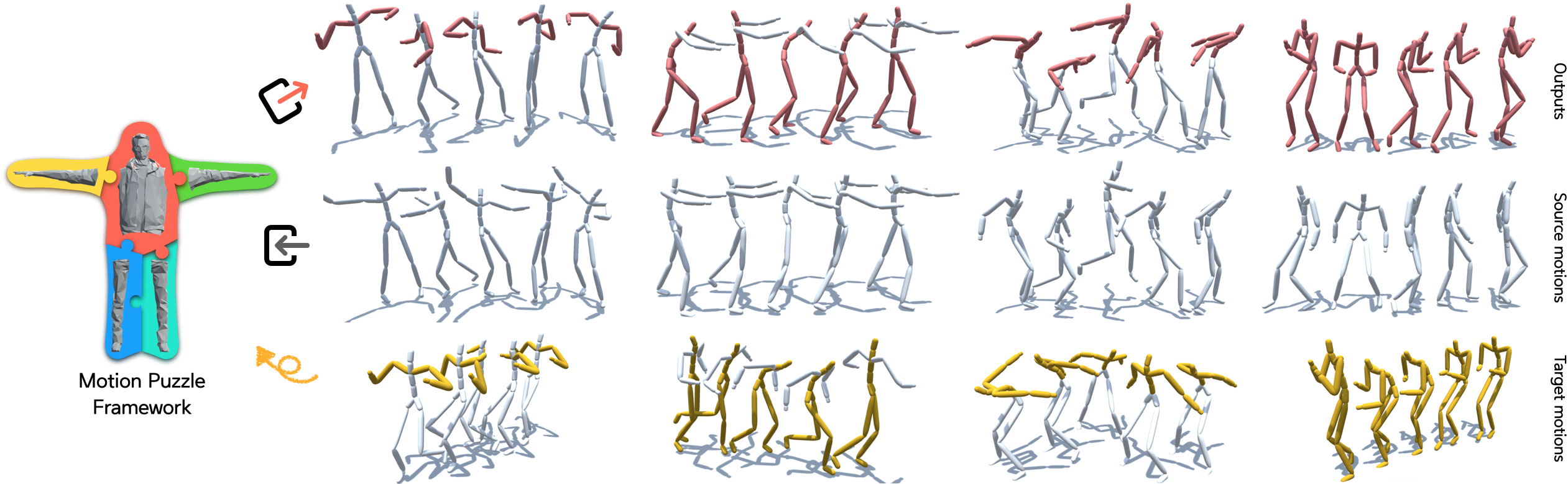}
  \caption{Our Motion Puzzle framework translates the motion of selected body parts in a source motion to a stylized output motion. It can also compose the motion style from multiple target motions for different body parts like a puzzle to stylize the full-body output motion while preserving the content of the source motion. From the left, this figure shows examples of transferring styles in the target motions for the spine and legs, spine and arms, arms only, and whole body.}
 \label{fig:teaser}
\end{teaserfigure}

\maketitle

\section{Introduction}
\label{sec:intro}
Style is an essential component in human movement as it reflects various aspects of people, such as personality, emotion, and intention. Thus the expressibility of style is crucial for animating human characters and avatars.
However, obtaining stylized motions is a challenging task: Since motion style is somewhat subtle and subjective, professional actors or animators are often necessary to capture or edit motions, and even with them many iterations may be required because the same style can be expressed in numerous ways.
Such difficulties make producing high-quality stylized motions costly and time-consuming.

An effective way of creating stylized motions is extracting a style component from example motion data (called target motion) and applying it to another motion (called source motion) containing desired content, for which researchers have achieved remarkable progress \cite{yumer2016spectral, holden2016deep, aberman2020unpaired, park2021diverse}.
Yet, previous methods still suffer from several limitations.

First, they require style labeling or motion pairing in training datasets for specific loss functions or computing the difference between motions, making it difficult to use many publicly available unlabeled motion datasets. 
%
Second, these methods only capture spatially and temporally averaged global style features of the target motion and do not capture local style features well. Thus, they are limited in extracting time-varying styles, styles exhibited by dynamic movements (e.g., fluttering or instantaneous falling).
Third, they can only modify the whole-body motion while, in practical scenarios, it is often needed to change the style of only some part of the body. 
For example, they cannot add a collapsing style (tripping motion) only to the spine and legs of a zombie walking to create a collapsing style zombie walking.
This operation is in analogy to the segment style transfer that changes styles of only the nose, eyes, or lips of the whole face in image transfer~\cite{shen2020interpreting}.

This paper presents a novel framework for motion style transfer with important improvements on these limitations. 
A unique advantage of our framework is that it can control the motion style of individual body parts (the arms, legs, and spine) while preserving the content of a given source motion; hence we name it \emph{Motion Puzzle}. The ability to stylize motion by body part can dramatically increase the range of expressible motions: Rather than simply transferring a single style to a whole body, by combining various style components applied to different body parts, or by applying a new style component to only a subset of body parts, it can synthesize a wide variety of stylized motions. To the best of our knowledge, our work is the first that achieves this per-body-part motion style transfer. 

Motion Puzzle framework consists of two encoders for extracting multi-level style features for the individual body parts from target motions and a content feature from a source motion, and one decoder to synthesize a stylized whole-body motion.
For per-body-part style transfer, we develop a novel Body Part Style Network (BP-StyleNet), which takes the human body part topology into account during the style injection process, and employ a skeleton aware graph convolutional network to extract and process features while preserving the structure of body parts.

Another major advantage of our framework is that it can transfer both global and local traits of motion style via two-step transfer modules: Body Part Adaptive Instance Normalization (BP-AdaIN) and Body Part Attention Network (BP-ATN) inside BP-StyleNet. BP-AdaIN module injects global style features into the content feature by applying AdaIN \cite{huang2017arbitrary} by body part. BP-ATN module transfers the locally semantic style features by constructing an attention map between content feature and style feature by body part. Especially, the time-varying motion style is captured via ATN module. For example, given a motion style of flapping with bent arms, our framework successfully extracts and transfers arm bending (global feature) and flapping (local feature).

Finally, our framework enables arbitrary (zero-shot) motion style transfer by learning to identify the style and content components of any motion. By using datasets with style labeling or motion pairing, previous methods learn style latent space labeled with general text, e.g., happy or old \cite{aberman2020unpaired, park2021diverse}. 
However, motion styles are often ambiguous to describe in text, especially if they are subtle, compositive or temporally changing. This results in significant variability of text labels and a weak association between given motion and labels \cite{HyeJi2019}. Therefore, while style labeling can be effective for basic categories such as emotion, it may hardly be extended to a broader range of styles that are challenging to label. With this awareness, we consider motion style as a specific way of movement to realize content (or task) in a given motion, and do not relate it with other styles in different motions using general text labels. Our approach allows using many publicly available motion datasets for training and transferring truly unseen arbitrary motion styles in test mode. 

Figure~\ref{fig:teaser} shows the idea of the Motion Puzzle framework: Given one source motion (middle row) for content and target body parts motions (yellow parts, bottom) for style, the Motion Puzzle framework translate the motion of selected body parts in a source motion to a stylized output motion (red parts, top).
In addition, our framework can be easily integrated with other motion generators. We demonstrate a real-time motion style transfer by integrating with an existing motion controller. 

The contributions of our work can be summarized as follows:
\begin{itemize}[itemsep=5.0pt]
\item We present the first motion style transfer method that can control the motion style of individual body parts.
\item  We propose a new two-step style adaptation network, BP-StyleNet, consisting of BP-AdaIN and BP-ATN, to transfer spatially and temporally global and local features of motion style, which presents a distinctive advantage: time-varying style features can be captured and translated.
\item Our architecture that considers the human skeleton structure and body parts makes arbitrary (zero-shot) style transfer possible without style labeling or motion pairing of training data. 
\end{itemize}
\section{Related work}
\label{sec:related}
In this section, we introduce previous research on stylizing human motion and some image style transfer methods that contributed much to motion stylization. 

\subsection{Arbitrary Image Style Transfer}
The methods for transferring arbitrary image styles aim at zero-shot learning to synthesize a content image adopting an arbitrary unlabeled style of another image. 
Gatys et al.~\shortcite{gatys2016image} showed that the style of an image can be represented by the Gram matrices, which compute correlations between the feature maps of pre-trained convolutional networks. Their method includes a computationally heavy optimization step, which can be replaced with faster feed-forward networks \cite{johnson2016perceptual}.
Huang and Belongie~\shortcite{huang2017arbitrary} introduced the adaptive instance normalization (AdaIN), which learns to adapt the affine parameters to the style input to enable transferring arbitrary styles.
Since the AdaIN simply modifies the mean and variance of the content image, it cannot sufficiently transfer the semantic features of style images.
Li et al.~\shortcite{li2017universal} transformed content features into a style feature space by applying whitening and coloring transformation (WCT) with aligning the covariance. WCT is computationally heavy to deal with high dimensional features and it does not perform well when content and style images are semantically different.
Sheng et al.~\shortcite{sheng2018avatar} introduced Avater-Net, a patch-based style decorator that can transfer style features to the semantically nearest content features. However, it cannot capture both global and local style patterns since the scale of patterns in the style images depends on the patch size.
Park et al.~\shortcite{park2019arbitrary} proposed a style-attention network (SANet) to integrate the local style features according to the semantically nearest content features for achieving good results with evident style patterns.

In our work, we develop a method that applies the AdaIN in a way that preserves the structure of human body to be able to transfer the global style feature of individual body parts. In addition, we leverage the attention network to transfer local style features to semantically matched content features.

\subsection{Motion Style Transfer}
Previous work of motion style transfer using early machine learning methods infers motion style with manually defined features. 
The work of \cite{hsu2005style} models the style difference of two motions with a linear time-invariant (LTI) system, which can transform the style of a new motion that is similar in content.
Ma et al.~\shortcite{ma2010modeling} modeled style variation of individual body parts with latent parameters, which are controlled by user-defined parameters through a Bayesian network.
Xia et al.~\shortcite{xia2015realtime} proposed a method to construct the mixtures of autoregressive (MAR) models online to represent style variations locally at the current pose and apply linear transformations to control the style. Another approach is to represent the motion style in the frequency domain using the Fourier Transform \cite{unuma1995fourier, yumer2016spectral}, which allows extracting style features from a small dataset without a need to conduct spatial matching between data.
These studies on motion style transfer are usually adequate only for a limited range of motions and may require special processing, such as time-warping and alignment, of the example motions or searching motion database.

Deep learning-based approaches have greatly improved the quality and the possible range of motion stylization.
Holden et al.~\shortcite{holden2016deep,holden2017fast} showed that the motion style can be transferred based on the Gram matrices \cite{gatys2016image} through motion editing in the latent space. 
Du et al.~\shortcite{du2019stylistic} presented a conditional Variational Autoencoder (CVAE) with the Gram matrices to construct the style-conditioned distribution for human motion.
These approaches require much computing time to extract style features through optimization and have a limitation in capturing complex or subtle motion features, making style transfer between motions with significantly different contents ineffective. 
Mason et al.~\shortcite{mason2018few} applied few-shot learning to synthesize stylized motions with limited motion data. However, similarly to \cite{du2019stylistic}, this method is confined to specific types of motion, such as locomotion.

Recently, Aberman et al.~\shortcite{aberman2020unpaired} applied the generative adversarial networks (GAN) based architecture with the AdaIN from FUNIT~\cite{liu2019few} model used in image style transfer.
They alleviated the restrictions on training data by allowing for training the networks with an unpaired dataset with style labels while preserving motion quality and efficiency. 
It, also notably, can transfer style from videos to 3D animations by learning a shared style embedding for both 3D and 2D joint positions.
Park et al.~\shortcite{park2021diverse} constructed a spatio-temporal graph to model a motion sequence, which improves style translation between significantly different actions. Their framework generates diverse stylization results for a target style by introducing a network that maps random noise to various style codes for the target style.
However, the above approaches have two limitations. First, they require labeled data for content or style to construct adversarial loss of GAN-based model, making zero-shot (arbitrary) motion style transfer impossible. They cannot use many publicly available, unlabeled motion datasets, while ease of obtaining a wide range of data is critical for a deep learning method to generalize to the data in the wild. 
Second, style transfer using only the AdaIN cannot capture time-varying motion style. Since the AdaIN simply modifies the mean and variance of the content features to translate the source motion, it captures temporally global features well but loses temporally local features.
Wen et al.~\shortcite{wen2021autoregressive} proposed a flow-based motion stylization method. 
Its probabilistic structure allows to generate various motions with a specific set of style, context and control, but it also suffers from capturing time-varying motion styles. 

In contrast, our framework achieves high-quality motion style transfer by using a novel model structure that does not require data labeling for adversarial loss.
Therefore, our framework can transfer truly unseen arbitrary motion styles. 
In addition, we solve the problem of capturing time-varying motion style by developing BP-StyleNet, which combines AdaIN-based module for the global style feature and the attention module for the local style feature.

\subsection{Skeleton-based Graph Networks}
\label{subsec:related_work_STGCN}

Since the human skeleton has specific connectivity, researchers have proposed network structures and associated operations that respect the skeleton topology to model correlation between joints.
Zhou et al.~\shortcite{zhou2014human} constructed a partition-based structure of human skeleton and used a graphical representation of the conditional dependency between joints to model variations of human motion.
Yan et al.~\shortcite{yan2018spatial} proposed a spatio-temporal graph convolution network (STGCN) to deal with skeleton motion data. STGCN consists of spatial edges that connect adjacent joints in the human body and temporal edges that connect the same joint in consecutive frames. 
The skeleton-aware graph structures have been successfully adopted in many applications~\cite{shi2019two, huang2020part, li2019actional, shi2019skeleton, si2018skeleton}.
Aberman et al.~\shortcite{aberman2020skeleton} proposed skeleton-aware operators for convolution, pooling and unpooling for motion retargeting between different homeomorphic skeletons. Park et al.~\shortcite{park2021diverse} used STGCN for motion style transfer.

Most previous work on deep learning-based motion style transfer~\cite{holden2016deep, holden2017fast, du2019stylistic, smith2019efficient, aberman2020unpaired} did not consider the hierarchical spatial and temporal structure of the human skeleton. Park et al.~\shortcite{park2021diverse} used STGCN to consider the skeletal joint structure during the convolution but did not take the human body part topology into account for the style injection.
In this work, we also use STGCN to represent spatial-temporal change of human motion but go a significant step further by proposing novel style adaptation networks, BP-StyleNet that consider human body parts during style injection process.
Thanks to this skeleton-aware transfer network, our framework can control style transfer per body part, which previous studies have not made possible.

\begin{figure*}[t]
  \centering
  \includegraphics[width=0.9\textwidth]{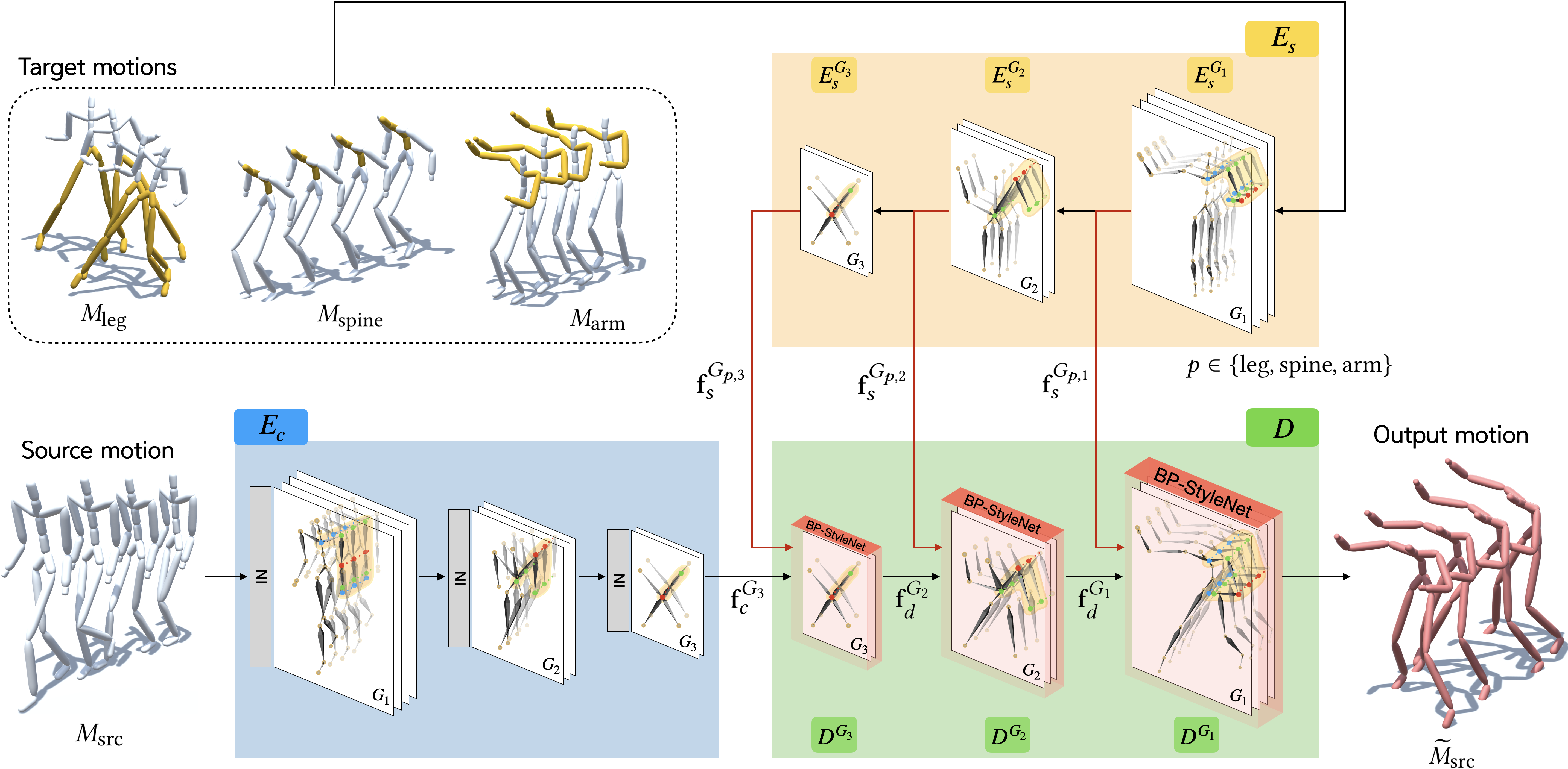}
  \caption{The overall network architecture of our Motion Puzzle framework. Our framework can transfer the styles of some body parts in the target motion to the corresponding body parts in the source motion. This figure illustrates a case where three target motions, each for the leg $M_\text{leg}$, spine $M_\text{spine}$, and arm $M_\text{arm}$, are used to stylize the source motion $M_\text{src}$ to generate the stylized output motion $\widetilde{M}_\text{src}$.
  Target motions are encoded into multi-level style features $\vf_{s}^{G_{p,i}}$ for the legs, arms and spine, respectively, through the style encoder $E_s$, and a source motion passes through the content encoder $E_c$ to extract a content feature $\vf_{c}^{G_{3}}$. The decoder $D$ of the framework receives the content feature and per-body-part multi-level style features, and progressively synthesizes a stylized whole-body intermediate features $\vf_{d}^{G_{i-1}}$ through a novel per-body-part style adaption network \emph{BP-StyleNet}.
  }
  \label{fig:framework}
\end{figure*}

\section{Motion data representation and processing}
\label{sec:data_processing}
Before discussing our framework in detail, we first explain the representation of motion data and the construction of the motion dataset.

We denote a human motion set by $\mathcal{M}$ and its subset of $T$ total number of frames by $\mathcal{M}_T \subset \mathcal{M}$, with $M_T$ being a random variable of $\mathcal{M}_T$.
We represent a motion of length $T$ as $M_T = [\mM_1, \ldots, \mM_t, \ldots, \mM_T]$, where $\mM_t$ denotes a pose feature matrix at frame $t$. Specifically, we define $\mM_t = [\vm_j^p, \vm_j^r, \dot{\vm}_j^p, \dot{\vr}^x, \dot{\vr}^z, \dot{\vr}^a]_{j=1}^{n_{joint}} = [\mM_{t, j}]_{j=1}^{n_{joint}}$, where $\vm_j^p \in \real^3$, $\vm_j^r \in \real^6$ and $\dot{\vm}_j^p \in \real^3$ are the local position, rotation and velocity of joint  $j$, expressed with respect to the character forward-facing direction in the same way as \cite{holden2016deep}. 
A joint rotation is represented by the 2-axis (forward and upward vectors) rotation matrix as in \cite{zhang2018mode}. 
Under the assumption that the root motion plays a critical role in motion style, each joint information is accompanied by the root velocity information: $\dot{\vr}^x \in \real$ and $\dot{\vr}^z \in \real$ are the root translational velocities in $X$ and $Z$ directions relative to the previous frame $t-1$, represented with respect to the current frame $t$, and $\dot{\vr}^a \in \real$ is the root angular velocity about the vertical (Y) axis.
Therefore, the dimension of our human motion feature with $T$ frames is 
$M_T \in \real^{T \times n_{joint} \times d_\text{joint}}$ and that of our pose feature matrix is $\mM_t \in \real^{n_{joint} \times d_\text{joint}}$, where $d_\text{joint}=15$ is the degrees of freedom of a joint. The data of our motion feature is arranged as a tensor of $T \times n_{joint} \times d_\text{joint}$ dimensions for our network architecture, whereas \cite{holden2016deep} used $T \times d_\text{body}$ matrices, where $d_\text{body}$ is the degrees of freedom of a pose. 

\subsection{Dataset Construction}
\label{par:clip}
We construct a motion dataset using the CMU motion data \cite{CMUdata}, which provides various motions with unlabeled styles.  All motion data are 60 fps and retargeted to a single 21-joint skeleton with the same skeleton topology as the CMU motion data. 
An additional dataset of \cite{xia2015realtime} is used as unseen test data for quantitative evaluation in Sec.~\ref{subsec:comparison}.
While motion data can have variable frame lengths at runtime, we train our networks with motion clips of 120 frames for convenience. For this, a long motion sequence is divided into 120-frame motion clips overlapping 60 frames with adjacent clips.

\paragraph{Data augmentation}
\label{par:aug}
To enrich the training dataset, we augment data in several ways.
First, we mirror the skeleton to double the number of original motion clips.
Second, we obtain additional data by changing the velocity of motion. For this, we randomly sample a sub-motion clip $M_{\Delta t}$ of length $\Delta t \in [\frac{T}{2},T]$ ($T=120$ in our experiment), and scale $M_{\Delta t}$ by a random factor $\gamma$, where $\gamma \in [1,2]$ if $\Delta t < \frac{3}{4}T$ or $\gamma \in [0.5,1]$ otherwise. The former effectively decreases the velocity of the motion, and the latter increases it. The scaled motion is either cut or padded to make 120 frames. We call this data augmentation method \emph{temporal random cropping}. We perform the temporal random cropping on input motion clip data with the rate of 0.2 during training.
Finally, we build a training dataset of about 120K motion clips of 120 frames in 60fps.
\section{Motion puzzle framework}
\label{sec:net}

Figure~\ref{fig:framework} shows the overall architecture of the Motion Puzzle framework. Inputs to our framework are one source motion for the content and multiple target motions, each for stylizing one or more body parts, and the output is a stylized whole-body motion.

To encode and synthesize complex human motions in which numerous joints move over time in spatially and temporally correlated manners, we use a spatial-temporal graph convolutional network as the basis of our framework and employ a graph pooling-unpooling method to keep the graph in accordance with human's body part structure (Sec.~\ref{subsec:STGCN}). 
Maximum five target motions are encoded into multi-level style features for the legs, arms and spine, respectively, through the style encoder (Sec.~\ref{subsec:sty_en}), and a source motion passes through the content encoder to extract a content feature (Sec.~\ref{subsec:con_en}). The decoder of the framework receives the content feature and per-part multi-level style features, and synthesizes a stylized whole-body motion. In this process, each level of style feature is adapted progressively into the spatial-temporal feature for the corresponding body part through a novel per-body-part style transfer network named \emph{BP-StyleNet} (Sec.~\ref{subsec:dec}).



\subsection{Skeleton based Graph Convolutional Networks}
\label{subsec:STGCN}

\paragraph{Skeletal graph construction}
\label{par:graph}
A human motion $M_T$ has a temporal relationship in the sequence of pose features $\mM_1, \ldots,\mM_T$, each of which shows hierarchical spatial relationship among the skeletal joints.
We use a spatial-temporal graph to model these characteristics. Specifically, we follow the graph structure proposed by \cite{yan2018spatial}, summarized below.

\begin{figure}[t]
  \begin{minipage}[c]{0.35\linewidth}
    \includegraphics[width=\linewidth]{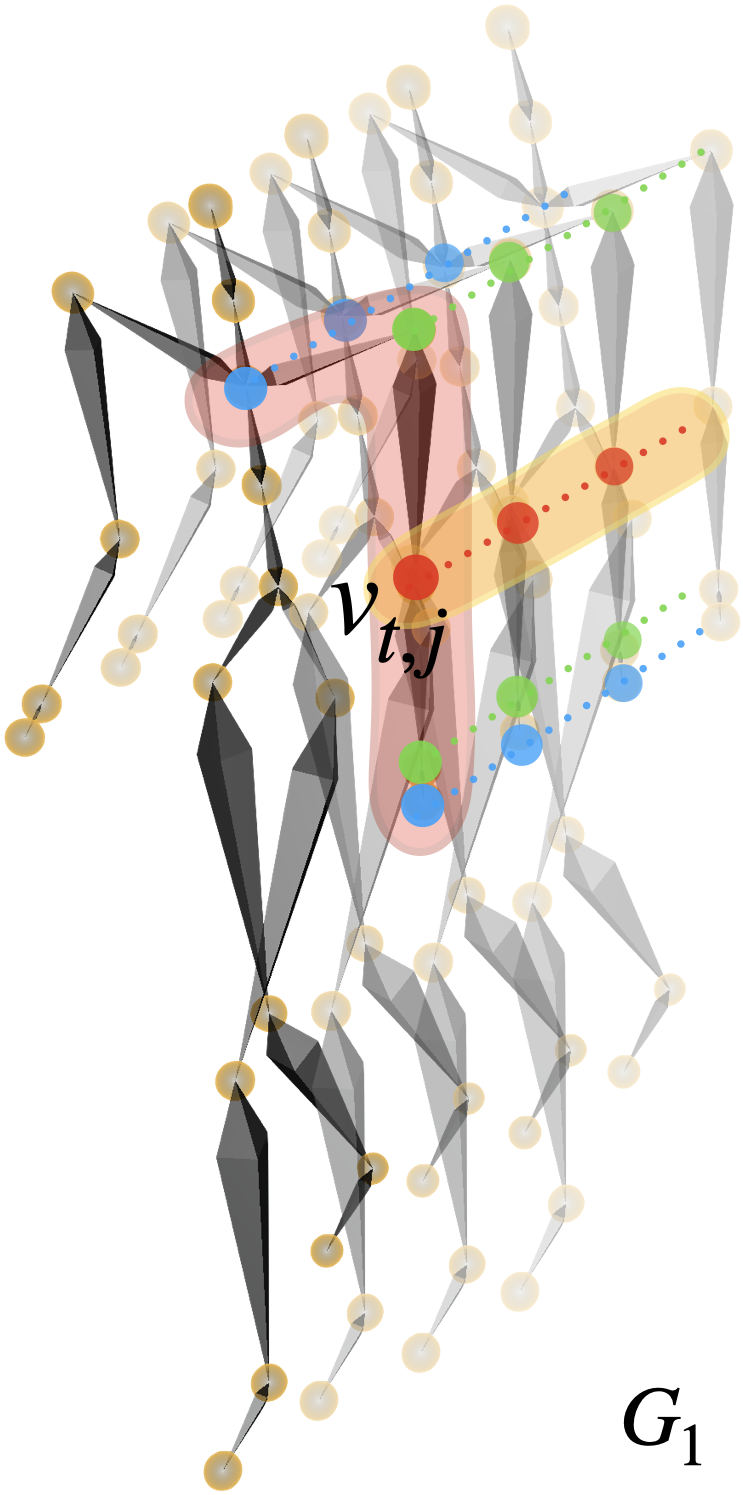}
  \end{minipage}\hfill
  \begin{minipage}[c]{0.6\linewidth}
    \caption{Illustration of the spatial-temporal graph.
  Vertices are denoted as yellow circles. Red, green and blue vertices denote 0, 1 and 2 edge distances from $v_{t,j}$.
  Black bones constitute the spatial edges, and dotted lines make the temporal edges. Note that this figure is only to visualize the connections of vertices from the perspective of the skeleton structure.  The features of each vertex in the graph are not directly related to vertex positions.
    } \label{fig:stgcn}
  \end{minipage}
\end{figure}

Figure~\ref{fig:stgcn} shows an example of a spatial-temporal graph $G_1=(V_1,E_1)$ featuring both the intra-body and inter-frame connections for a skeleton sequence with $n_\text{joint}$ joints and $T$ frames.
The vertex set $V_1 = \{v_{t,j}|t=1,\ldots,T \; \text{and} \; j=1,\ldots,n_\text{joint}\}$ corresponds to all the joints in the skeleton sequence.
The edge set $E_1$ consists of two kinds: Spatial edges $E_B = \{v_{t,j}v_{t,i}|(j,i) \in S \}$, where $S$ is a set of connected  joint pairs, connect adjacent joints at each frame $t$, and the temporal edges $E_F = \{v_{t,i}v_{t-1,i}\}$ connect the vertices for the same joint in adjacent time frames.

\paragraph{Graph convolution}
We describe the spatial-temporal graph convolution using the case of vertex $v_{t,j}$ in graph $G_1$ shown in Fig.~\ref{fig:stgcn}.
The spatial graph convolution at frame $t$ is formulated as follows \cite{yan2018spatial}:
\begin{equation} \label{eq:stgcn}
    f_{out}(v_{t,j}) = \sum_{v_{t,i} \in \mathcal{B}_{t,j}} \frac{1}{Z_{t,j}}f_{in}(v_{t,i}) \cdot \vw(d(v_{t,i}, v_{t,j})),
\end{equation}
where $f$ denotes the feature map, e.g., $f(v_{t,j})=\mM_{t,j}$ for graph $G_1$ if no operation has been performed yet, and $\mathcal{B}_{t,j}$ is the neighboring vertices for convolving $v_{t,j}$, defined as the $K$ edge-distance neighbor vertices $v_i$ from the target vertex $v_j$:
\begin{equation} \label{eq:sampling}
    \mathcal{B}_{t,j} = \{v_{t,i}|d(v_{t,i}, v_{t,j}) \leq K \},
\end{equation}
where $K=2$ for $G_1$.

The weight function $\vw$ provides a weight vector for the convolution, and we choose to model it as a function of edge distance $d(v_{t,i}, v_{t,j})$. That is, the vertices of the same color in Fig.~\ref{fig:stgcn} are given the same weight vector $\vw$ to convolve $v_{t,j}$.
Note that while the number of convolution weight vectors is fixed, the number of vertices in $\mathcal{B}_{t,j}$ varies, of which effect is normalized by dividing by the number $Z_{t,j}$ of vertices of the same edge distance.
The convolution in the temporal dimension is conducted straightforwardly as regular convolution because the structure of the temporal edge is fixed. 
Graph convolutions on $G_2$ and $G_3$ are performed in the same way with $K=1$.

\paragraph{Graph Pooling and Unpooling}
\label{subsec:graph_pooling}
Recent image transfer methods~\cite{karras2019style, liu2019few} show remarkable performance in translating images by gradually increasing resolution and adding details while unpooling (upsampling) high-level features and injecting style features.
Following this idea, we incorporate the pooling and unpooling method into our spatial-temporal graph to handle and extract features from the low (local) to high (global) level.

\begin{figure}[t]
  \centering
  \includegraphics[width=\linewidth]{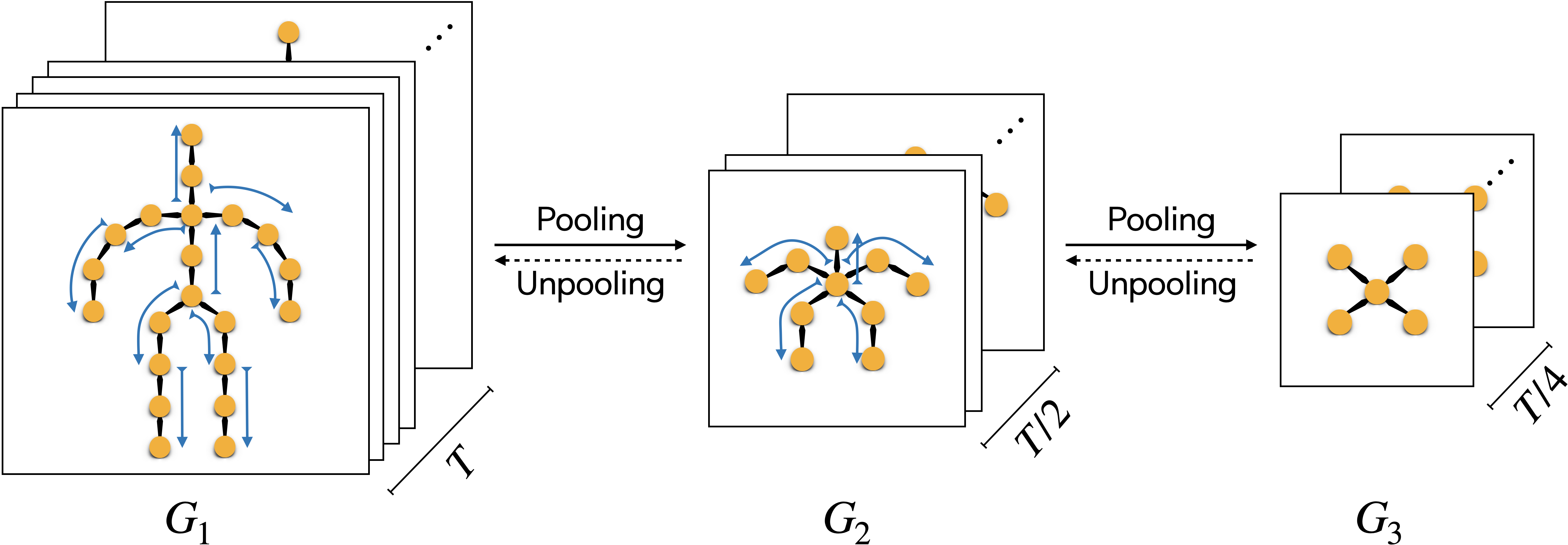}
  \caption{Body part preserving pooling and unpooling of the spatial-temporal graph. Blue arrows indicate the vertices used by each pooling kernel along the kinematics chain directions.}
  \label{fig:pooling}
\end{figure}

We use a standard average pooling method for the temporal dimension of the graph as it is uniformly structured with vertices sequentially connected along time.
However, as its spatial dimension is non-uniformly structured, standard pooling methods are not applicable. For this, several methods~\cite{yan2019convolutional, aberman2020skeleton}  exist to preserve the skeletal structure for motion retargeting and synthesis. We devise a similar graph pooling method (Fig.~\ref{fig:pooling}), with a difference being that our pooling operator (blue arrow) averages consecutive vertices along the kinematic chain only within the same body part to retain the individuality of the body parts. 
As a result of pooling, our framework uses graph structures $G_i=(V_i, E_i)$ in varying resolutions. Specifically, $|V_1|=21\times T$, $|V_2|=10\times \frac{T}{2}$ and $|V_3|=5\times \frac{T}{4}$. Note that each vertex $v_{t,j}$ in $G_3$ corresponds to each of the 5 body parts. The edges $E_2$ and $E_3$ are constructed similarly as $E_1$.

Unpooling is conducted in the opposite direction of the pooling. We use the nearest interpolation for upscaling the temporal dimension. For the spatial dimension, we unpool the joint dimension by mapping the pooled vertices to the previous skeleton structure. 

\subsection{Style Encoder}
\label{subsec:sty_en}
We develop our style encoder to extract the style features in multi-level for gradual translation.
The style encoder $E_s$ is a concatenation of multi-level encoding blocks $E_s^{G_i},\,i\in\{1,2,3\}$ corresponding to graph $G_i$, each of which consists of a STGCN layer followed by a graph pooling layer.
Each encoding block progressively extracts the intermediate style feature $\vf_{s}^{G_i} = E_s^{G_i}(\vf_{s}^{G_{i-1}})$.
As a result, given a target motion $M_p$, the style encoder $E_s$ extracts multi-level style features.
\begin{equation} \label{eq:enc_style}
    [\vf_{s}^{G_{p,i}}]_{i=1}^3 = E_s(M_p),
\end{equation}
where $G_{p,i}$ is the graph from $M_p$, $\vf_{s}^{G_{p,1}} \in \real^{21 \times T_\text{p} \times C}$, $\vf_{s}^{G_{p,2}} \in \real^{10 \times \frac{T_\text{p}}{2} \times 2C}$ and $\vf_{s}^{G_{p,3}} \in \real^{5 \times \frac{T_\text{p}}{4} \times 4C}$ with $T_\text{p}$ being the frame number of $M_p$ and $C$ being the feature dimension. These style features are used to transfer styles progressively during the decoding process.

Each level of feature $\vf_{s}^{G_{p,i}}$ can be divided into five part-features that correspond to the five body parts thanks to STGCN and the part-preserving pooling scheme. 
\begin{equation}
\vf_{s}^{G_{p,i}} = [\vf_{s}^{LL_{p,i}}, \vf_{s}^{RL_{p,i}}, \vf_{s}^{SP_{p,i}}, \vf_{s}^{LA_{p,i}}, \vf_{s}^{RA_{p,i}} ],
\end{equation}
where $LL, RL, SP, LA,$ and $RA$ denote subset vertices of graph $G$ corresponding to the left leg, right leg, spine, left arm, and right arm, respectively. 
This structure-aware style transfer is reasonable for the graph convolution-based approach and leads to a better stylization performance, as will be shown in Sec.~\ref{subsec:comparison}.

Leveraging this division, from one target motion, we can select only a subset of features corresponding to the target body parts to apply the encoded style, and for other body parts, we can use a subset of encoded style features of other target motion. 
Thus, the whole set of style features can be made by combining these part-features from up to 5 different motions $p,\ldots,t$.
\begin{equation} \label{eq:bdypart_features}
\vf_{s}^{G_{i}} = [\vf_{s}^{LL_{p,i}}, \vf_{s}^{RL_{q,i}}, \vf_{s}^{SP_{r,i}}, \vf_{s}^{LA_{s,i}}, \vf_{s}^{RA_{t,i}} ].
\end{equation}
Note that each part-feature is allowed to have different temporal lengths.
This approach gives much freedom for controlling styles. We can use only one target motion for the whole body parts ($p=\ldots=t$) or three motions for the arm, spine, and leg ($p=q$ and $s=t$), both of which can be general use cases of our method. For those parts that we want to preserve the original style of the source motion, the style feature of the source motion can be used.
Hereafter, we will omit motion IDs $p,\ldots,t$ but assume that $\vf_{s}^{G_{i}}$ is made by one or more motions.
Figure~\ref{fig:framework} shows a case where three target motions $M_\text{leg}, M_\text{spine}$, and $M_\text{arm}$ are used for transferring styles to the legs, spine, and arms. 


\subsection{Content Encoder}
\label{subsec:con_en}
The content encoder $E_c$ has a similar structure as the style encoder.
The differences are that every STGCN layer is preceded by Instance Normalization (IN) to normalize feature statistics and remove style variations, and only the final output $\vf_c^{G_3}$ is used for the content feature.  
As a result, the content encoder extracts the style-invariant latent representation for motion.
Given a source motion $M_\text{src}$, the encoding process (blue part in Fig.~\ref{fig:framework}) is written as:
\begin{equation} \label{eq:enc_content}
    \vf_c^{G_3} = E_c(M_\text{src}),
\end{equation}
which maps the low-level joint feature $M_\text{src} \in \real^{21 \times T_c \times d_\text{joint}}$ to a high-level body part content feature $\vf_c^{G_3} \in \real^{5 \times \frac{T_c}{4} \times 4C}$. 

Same as the style feature, the content feature has the graph structure in accordance with human's five body parts, which plays a key role not only in better reconstructing human motion but also in performing per-body-part style transfer during the decoding step.

\subsection{Decoder}
\label{subsec:dec}
\begin{figure}[t]
  \centering
  \includegraphics[width=0.7\linewidth]{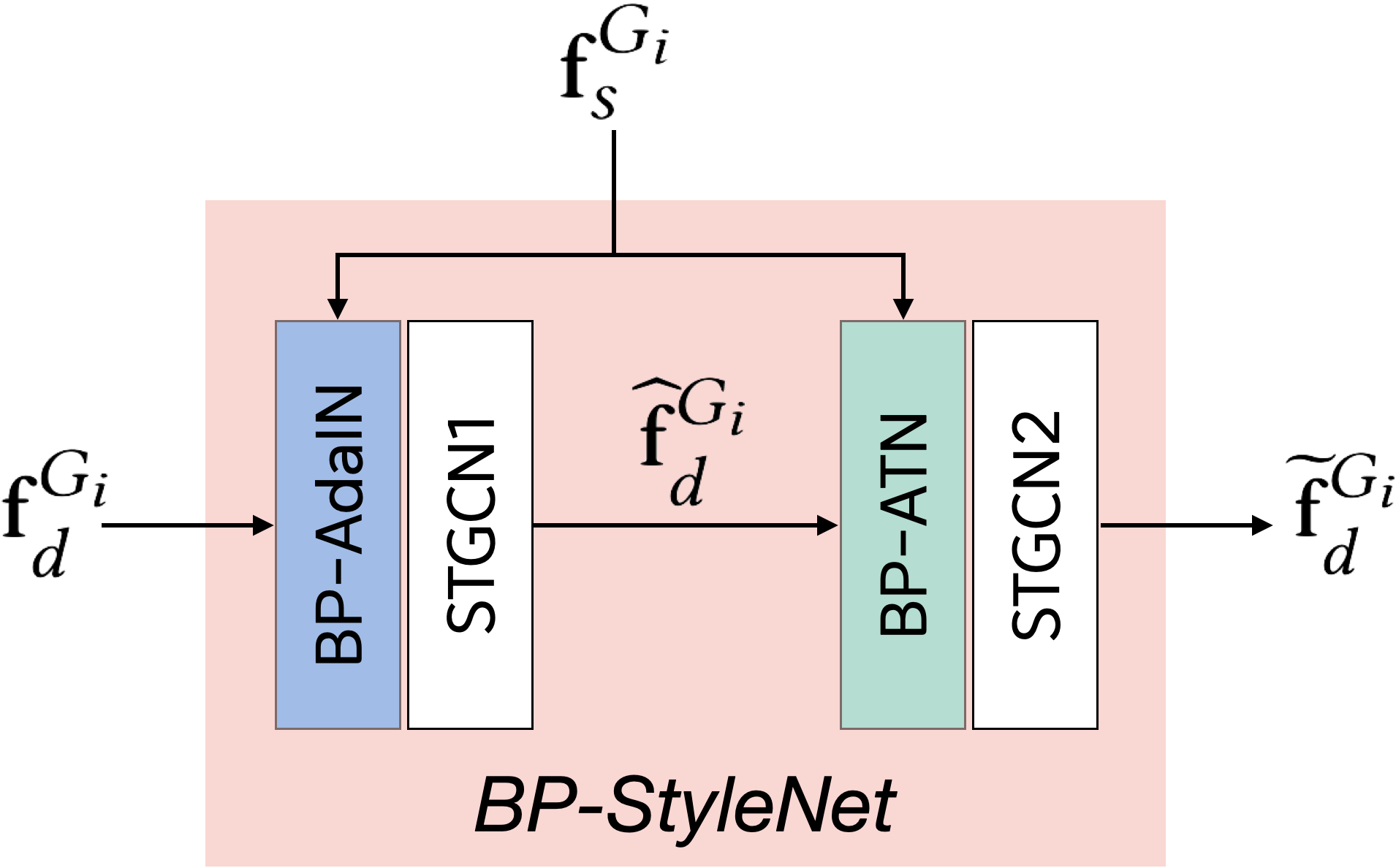}
  \caption{BP-StyleNet on $G_i$ level. It transfers the global and local characteristics of a style feature $\vf_s^{G_i}$ to $\vf_{d}^{G_i}$ via  two-step transfer modules, BP-AdaIN and BP-ATN.}
  \label{fig:BP_StyleNet}
\end{figure}

The decoder $D$ (green part in Fig.~\ref{fig:framework}) transforms the content feature $\vf_c^{G_3}$ into an output motion $\widetilde{M}_\text{src}$ using the multi-level target style features $[\vf_s^{G_i}]_{i=1}^3$.
\begin{equation} \label{eq:translator}
    \widetilde{M}_\text{src} = D(\vf_c^{G_3},\,[\vf_s^{G_i}]_{i=1}^3)
\end{equation}
The decoder network takes an inverse form of the encoder networks.
Similarly to the multi-level style adaptation module for image style transfer~\cite{sheng2018avatar}, we progressively generate the intermediate features $\vf_{d}^{G_{i-1}} = D^{G_i}(\vf_{d}^{G_i})$ starting from $\vf_{d}^{G_3} (=\vf_{c}^{G_3})$, with $\dim({\vf_{d}^{G_{i}}})=\dim({\vf_{c}^{G_{i}}})$.
Each level of decoding block $D^{G_{i}}$ is composed of our novel style adaption network \emph{BP-StyleNet} and an unpooling layer. 

\paragraph{\textbf{BP-StyleNet.}}
In order to transfer motion styles, we propose \emph{BP-StyleNet}.
As shown in Fig.~\ref{fig:BP_StyleNet}, it translates a decoded content feature $\vf_{d}^{G_i}$ into feature $\widetilde{\vf}_{d}^{G_i}$ with style feature  $\vf_s^{G_i}$ via two-step transfer modules, namely BP-AdaIN and BP-ATM.
In the first step, BP-AdaIN (Adaptive Instance Normalization) transfers the global statistics of style feature to generate $\widehat{\vf}_{d}^{G_i}$. Next, BP-ATN (attention network) transfers the local statistics of style feature to create $\widetilde{\vf}_{d}^{G_i}$, which now reflects both the global and local traits of style feature. 

\begin{figure}[t]
  \centering
  \includegraphics[width=0.7\linewidth]{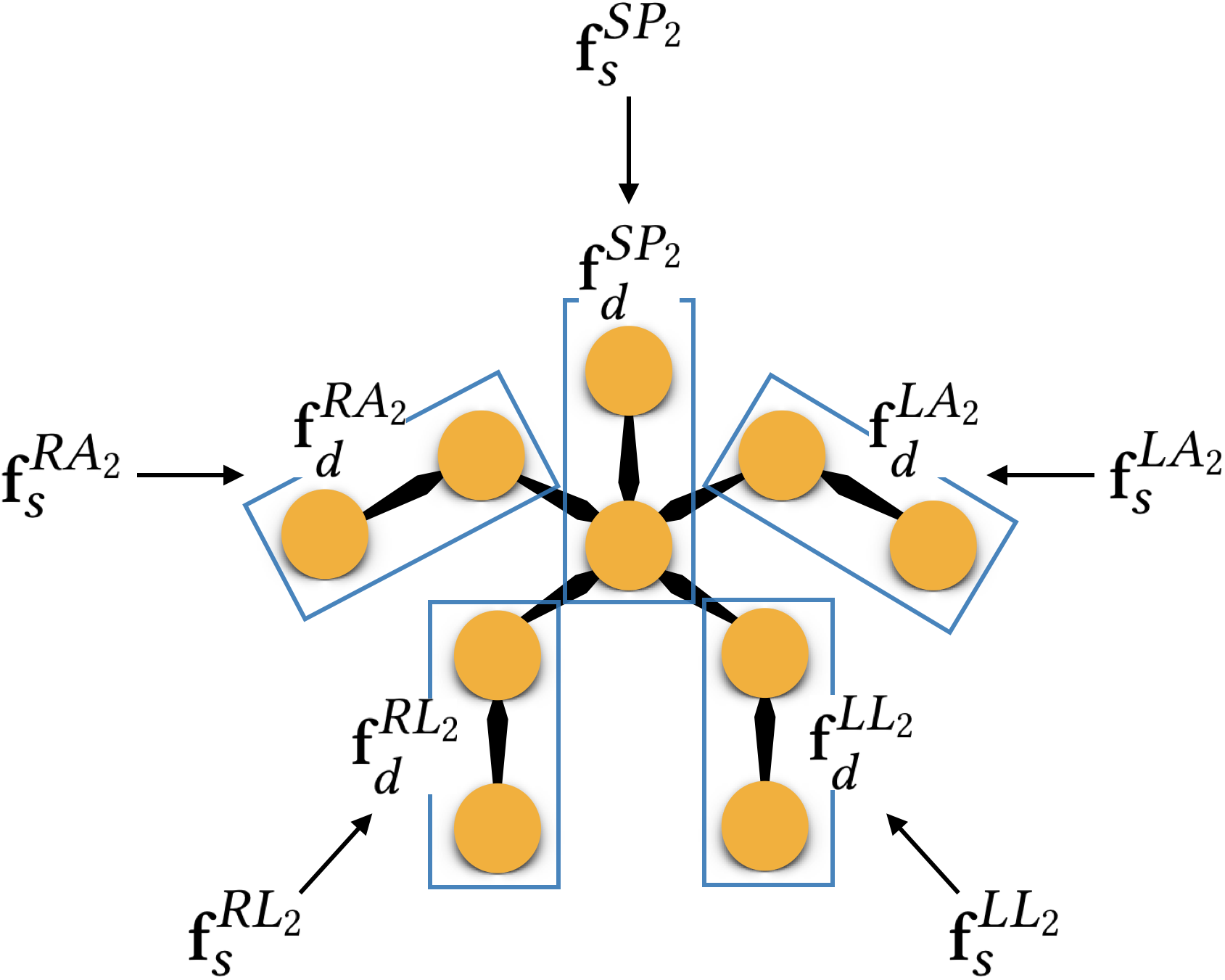}
  \caption{Example of applying BP-AdaIN (blue boxes) on $G_2$ level. The BP-AdaIN module injects style feature $\vf_s^{G_2}$ into $\vf_{d}^{G_2}$ by body part.}
  \label{fig:BP_module}
\end{figure}

The BP-AdaIN module takes $\vf_{d}^{G_i}$ as input and injects style feature $\vf_s^{G_i}$ by applying AdaIN by body part: 
\begin{equation} \label{eq:BodyPartAdaIN}
\text{BP-AdaIN}(\vf_{d}^{G_i}, \vf_{s}^{G_i}) = [\text{AdaIN}(\vf_{d}^{LL_i}, \vf_{s}^{LL_i}),\ldots,\text{AdaIN}(\vf_{d}^{RA_i}, \vf_{s}^{RA_i})].
\end{equation}
Figure~\ref{fig:BP_module} shows an example of how BP-AdaIN injects style features into each body part in $G_2$.
The style injection process of AdaIN is conducted as:
\begin{equation} \label{eq:AdaIN}
    \text{AdaIN}(\vf_{d}^{P_i}, \vf_{s}^{P_i}) = 
    \gamma^{P_i}(\vf_{s}^{P_i}) \left( \frac{\vf_{d}^{P_i} - \mu(\vf_{d}^{P_i})}{\sigma(\vf_{d}^{P_i})} \right) + \beta^{P_i}(\vf_{s}^{P_i}), 
\end{equation}
where $P\in\{LL, RL, SP, LA, RA\}$, $\mu$ and $\sigma$ are the channel-wise mean and variance, respectively.
AdaIN scales the normalized $\vf_{d}^{P_i}$ with a learned affine transformation with scales $\gamma^{P_i}$ and biases $\beta^{P_i}$ generated by $\vf_{s}^{P_i}$.
Finally, we feed output feature into STGCN1 layer and generate the first stage of the stylized feature.
\begin{equation}
    \widehat{\vf}_d^{G_i}=\text{STGCN1}(\text{BP-AdaIN}(\vf_{d}^{G_i})).
\end{equation}

In the second step, the BP-ATN module transfers the locally semantic style features of $\vf_{s}^{G_i}$ via constructing an attention map with  $\widehat{\vf}_d^{G_i}$ by body part.
The BP-ATN module was inspired by ~\cite{wang2018non, park2019arbitrary, fu2019dual}, which use attention block for action recognition, image segmentation, and image style transfer.
We feed the globally-stylized decoded feature $\widehat{\vf}_d^{G_i}$ and style feature $\vf_s^{G_i}$ to the BP-ATN module that maps the correspondences between the part-features of the same body part.

\begin{equation} \label{eq:BodyPartATN}
\text{BP-ATN}(\widehat{\vf}_{d}^{G_i}, \vf_s^{G_i}) = [\text{ATN}(\widehat{\vf}_{d}^{LL_i}, \vf_{s}^{LL_i}),\ldots,\text{ATN}(\widehat{\vf}_{d}^{RA_i}, \vf_{s}^{RA_i})].
\end{equation}

\begin{figure}[t]
  \centering
  \includegraphics[width=0.9\linewidth]{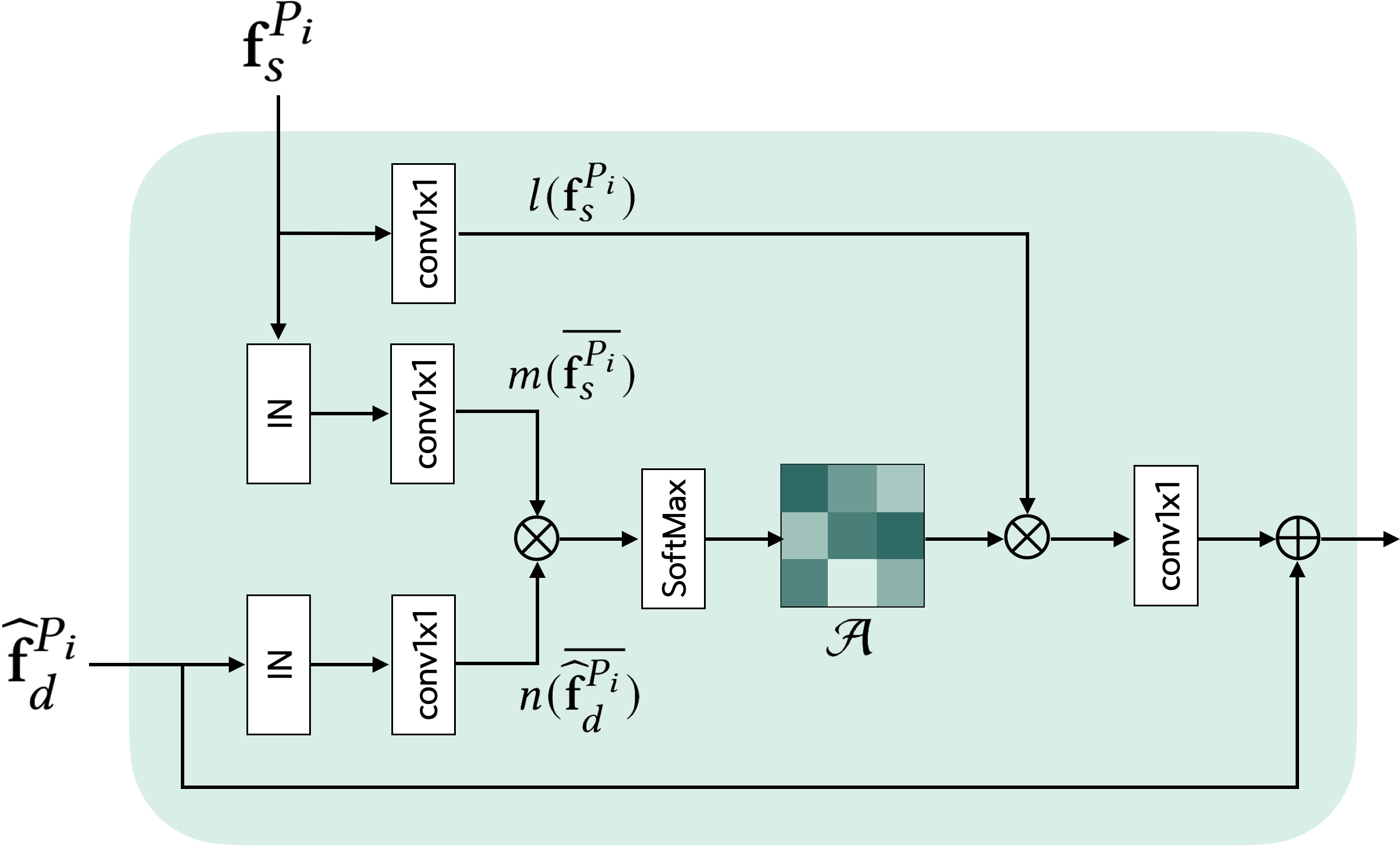}
  \caption{Applying BP-ATN module by body part $P_i$ of $G_i$ level. The BP-ATN module transfers the locally semantic style features of $\vf_{s}^{P_i}$ via constructing an attention map with $\widehat{\vf}_d^{P_i}$. }
  \label{fig:attention_module}
\end{figure}

Figure~\ref{fig:attention_module} illustrates the ATN module. It channel-wise normalizes the given style feature $\vf_{s}^{P_i}$ and the decoded content feature $\widehat{\vf}_{d}^{P_i}$ to make $\overline{{\vf}_{s}^{P_i}}$ and $\overline{\widehat{\vf}_{d}^{P_i}}$, and maps them into new feature spaces $m$ and $n$ by convolution layers. 
Then we reshape the mapped features to $m(\overline{{\vf}_{s}^{P_i}}) \in \real^{C_i \times |V_i|_s}$ and $n(\overline{\widehat{\vf}_{d}^{P_i}}) \in \real^{|V_i|_d \times C_i}$, where $|V_i|$ is vertex cardinality of part-features and $C_i$ is the feature dimension of $G_i$. 
After that, we perform a matrix multiplication between $m$ and $n$, and apply a softmax layer to construct an attention map $\mathcal{A} \in \real^{|V_i|_s \times |V_i|_d}$:
\begin{equation} \label{eq:attention_map}
    \mathcal{A}_{\beta,\alpha} = \frac{\exp(m(\overline{{\vf}_{s}^{P_i}})_\alpha \cdot n(\overline{\widehat{\vf}_{d}^{P_i}})_\beta)}
    {\sum_{\alpha=1}^{|V_i|_s}\exp(m(\overline{{\vf}_{s}^{P_i}})_\alpha \cdot n(\overline{\widehat{\vf}_{d}^{P_i}})_\beta)},
\end{equation}
where $\mathcal{A}_{\beta,\alpha}$ measures the similarity between the $\alpha$-th vertex of the style feature and the $\beta$-th vertex of the decoded feature. The more similar feature representation of the two vertices signifies the greater spatial-temporal correlation between them.
Meanwhile, we feed $\vf_{s}^{P_i}$ to another convolution layer and reshape it to generate a new feature, $l(\vf_{s}^{P_i}) \in \real^{|V_i|_s \times C_i}$. 
Then, we perform a matrix multiplication between $\mathcal{A}$ and $l(\vf_{s, P}^{P_i})$ to adjust the attention map to the style feature. After that, we apply a convolution layer to the result and add to $\widehat{\vf}_{d}^{P_i}$ element-wise: 
\begin{equation}
    \text{ATN}(\widehat{\vf}_{d}^{P_i}, \vf_{s}^{P_i}) = \mathcal{A}^T \otimes l(\vf_{s}^{P_i}) \oplus \widehat{\vf}_{d}^{P_i},
\end{equation}
where $\otimes$ denotes matrix multiplication and $\oplus$ element-wise addition.
Finally, we feed the output feature into STGCN2 layer and generate the second step of the stylized feature $\widetilde{\vf}_d^{G_i}$;
\begin{equation}
    \widetilde{\vf}_d^{G_i} = \text{STGCN2}(\text{BP-ATN}(\widehat{\vf}_{d}^{G_i}, \vf_s^{G_i})).
\end{equation}

Implementation details for the whole architecture are provided in Appendix~\ref{Appendix:architecture}.

\section{Training the Motion Puzzle framework}
\label{sec:training}
Given a source motion $M_\text{src}\ \in \mathcal{M}$ and a target motion $M_\text{tar} \in \mathcal{M}$, we train the entire networks end-to-end by using the following techniques and loss terms.

\begin{description}[leftmargin=0pt, itemsep=5.0pt, labelindent=\parindent, listparindent=\parindent]
\item[Mixing style code by body part.]
A naive way of preparing the training data for our per-body-part style transfer would be to construct a batch of motions by combining one source and many target motions, but it requires too much memory. To avoid this, we devise a per-body-part style mixing scheme following the idea of \cite{karras2019style} to achieve a similar effect with only one source and one target motion.
The idea is, instead of explicitly preparing target style features, a style feature ${\vf_s^{G_i}}'$ from a target motion and a style feature $\vf_s^{G_i}$ from a source motion are mixed with a predetermined probability to apply the target and source style codes to some randomly selected body parts, as implemented in Algorithm~\ref{alg:style_mixing}.

\begin{algorithm}[t]
\DontPrintSemicolon
\KwIn{Style feature set $\vf_s^{G_i}$ from source motion and ${\vf_s^{G_i}}'$ from target motion}
\KwOut{Mixed style feature set $\vf_\text{mix}^{G_i}$}
\tcc{$\gamma$: predefined probability of style mixing}
\eIf{$Rand([0,1])<\gamma$}
    {
        \tcc{Number of parts to apply $\vf_s^{G_i}$} 
        $n_\text{switch} = RandInt([1,5])$\;
        \tcc{Randomly selected $n_\text{switch}$ number of body parts} 
        $idx_\text{switch} = Sample(idx_\text{all}, n_\text{switch})$\;
        \tcc{Mix style codes} 
        $\vf_\text{mix}^{G_i}[idx_\text{switch}] = \vf_s^{G_i}$\;
        $\vf_\text{mix}^{G_i}[idx_\text{all}-idx_\text{switch}] ={\vf_s^{G_i}}'$\;
    }
    {
        \tcc{Only use style code ${\vf_s^{G_i}}'$} 
        $\vf_\text{mix}^{G_i}[idx_\text{all}] ={\vf_s^{G_i}}' $\;
    }
\caption{Style mixing by body part.}
\label{alg:style_mixing}
\end{algorithm}

\item[Motion reconstruction.]
By using the same motion for both the source and target motions, the reconstruction loss enforces the translator to generate  output motions identical to the input motions.
\begin{equation} \label{eq:rec}
    \begin{aligned}
    \mathcal{L}_{rec} &= \mathbb{E}_{M_\text{src}}[\| \widetilde{M}_\text{src}^\text{rec}- M_\text{src}\|_1] + \mathbb{E}_{M_\text{tar}}[\|\widetilde{M}_\text{tar}^\text{rec} - M_\text{tar}\|_1], \\
    \widetilde{M}_\text{src}^\text{rec} &= D(\vf_c^{G_3},\,[\vf_s^{G_i}]) \text{ (reconstructed source motion)},\\
    \widetilde{M}_\text{tar}^\text{rec} &=
D({\vf_c^{G_3}}',\,[{\vf_s^{G_i}}']) \text{ (reconstructed target motion)},
    \end{aligned}
\end{equation}
where $\vf_c^{G_3} = E_c(M_\text{src})$, $[\vf_s^{G_i}]=E_s(M_\text{src})$ while ${\vf_c^{G_3}}'$ and $[{\vf_s^{G_i}}']$ are from the target motion ($[\cdot]_{i=1}^3$ is written as $[\cdot]$ for brevity). This loss term helps our encoder-decoder framework achieve the identity map. 

\item[Preserving content and style features.]
To guarantee that the translated motion preserves the style-invariant characteristics of its input source motion $M_\text{src}$ and also preserves the content-invariant characteristics of its input target motion $M_\text{tar}$, we employ the cycle consistency loss ~\cite{zhu2017unpaired, lee2018diverse, choi2020stargan}.
\begin{equation} \label{eq:cyc}
    \begin{aligned}
    \mathcal{L}_{cyc} &= \mathbb{E}_{M_\text{src}, M_\text{tar}}[\| \widetilde{M}_\text{src}^\text{cyc} -  M_\text{src}\|_1 + \| \widetilde{M}_\text{tar}^\text{cyc} -  M_\text{tar}\|_1], \\
    \widetilde{M}_\text{src}^\text{cyc} &= D(E_c(D(\vf_c^{G_3}, [\vf_\text{mix}^{G_i}])), [\vf_s^{G_i}]), \\
    \widetilde{M}_\text{tar}^\text{cyc} &= D({\vf_c^{G_3}}', E_s(D(\vf_c^{G_3}, [{\vf_s^{G_i}}']))),
    \end{aligned}    
\end{equation}
where $[\vf_\text{mix}^{G_i}]$ is a mixed style feature set from $[\vf_s^{G_i}]$ and $[{\vf_s^{G_i}}']$ (Alg.~\ref{alg:style_mixing}). 

\item[Preserving root motion.]
We encourage the output motion to preserve the linear and angular velocities of the root of the source motion $M_\text{src}$ because the root motion plays a critical role in motion content. The preservation of the root motion gives an additional benefit of reducing foot sliding and improving temporal coherence of output motion. To this end, we employ a root loss as follows:
\begin{equation} \label{eq:loss_root}
    \mathcal{L}_{root} = \mathbb{E}_{M_\text{src}, M_\text{tar}}[\| RV (D(\vf_c^{G_3}, [\vf_\text{mix}^{G_i}])) - RV\left(M_\text{src}\right) \|_1],
\end{equation}
where $RV(M)$ extracts the root velocity terms $[\dot{\vr}^x, \dot{\vr}^z, \dot{\vr}^a]$ from a motion.

\item[Smoothness.]
To ensure that the output pose changes smoothly in time, we add a smoothness term between temporally adjacent pose features.
\begin{equation} \label{eq:loss_sm}
    \mathcal{L}_{sm}(\widetilde{M}, M) = \mathbb{E}_{M}[\| V(\widetilde{M}) - V(M) \|_1]
\end{equation}
\begin{equation} \label{eq:loss_sm_full}
    \begin{aligned}
    \mathcal{L}_{sm} &= \mathcal{L}_{sm}(\widetilde{M}_\text{src}^\text{rec}, M_\text{src}) + 
    \mathcal{L}_{sm}(\widetilde{M}_\text{tar}^\text{rec}, M_\text{tar}) \\
    &+ \mathcal{L}_{sm}(\widetilde{M}_\text{tar}^\text{cyc}, M_\text{src}) +
    \mathcal{L}_{sm}(\widetilde{M}_\text{tar}^\text{cyc}, M_\text{tar})
    \end{aligned}
\end{equation}
\end{description}
where $V(M)$ is the sequence of pose feature differences between adjacent frames.

The total objective function of the Motion Puzzle framework is thus:
\begin{equation} \label{eq:loss_total}
    \min_{E_c,E_s,D} \; \mathcal{L}_{rec} + \lambda_{cyc}\mathcal{L}_{cyc}
    + \lambda_{root}\mathcal{L}_{root} + \lambda_{sm}\mathcal{L}_{sm},
\end{equation}
where $\lambda_{cyc}$, $\lambda_{root}$ and  $\lambda_{sm}$ are hyperparameters for each loss term. Please see training details in Appendix~\ref{Appendix:training}.

\section{Discussion on Content Feature and Attention Map}
\label{sec:discussion}
\begin{figure}[t]
  \centering
  \includegraphics[width=0.8\linewidth]{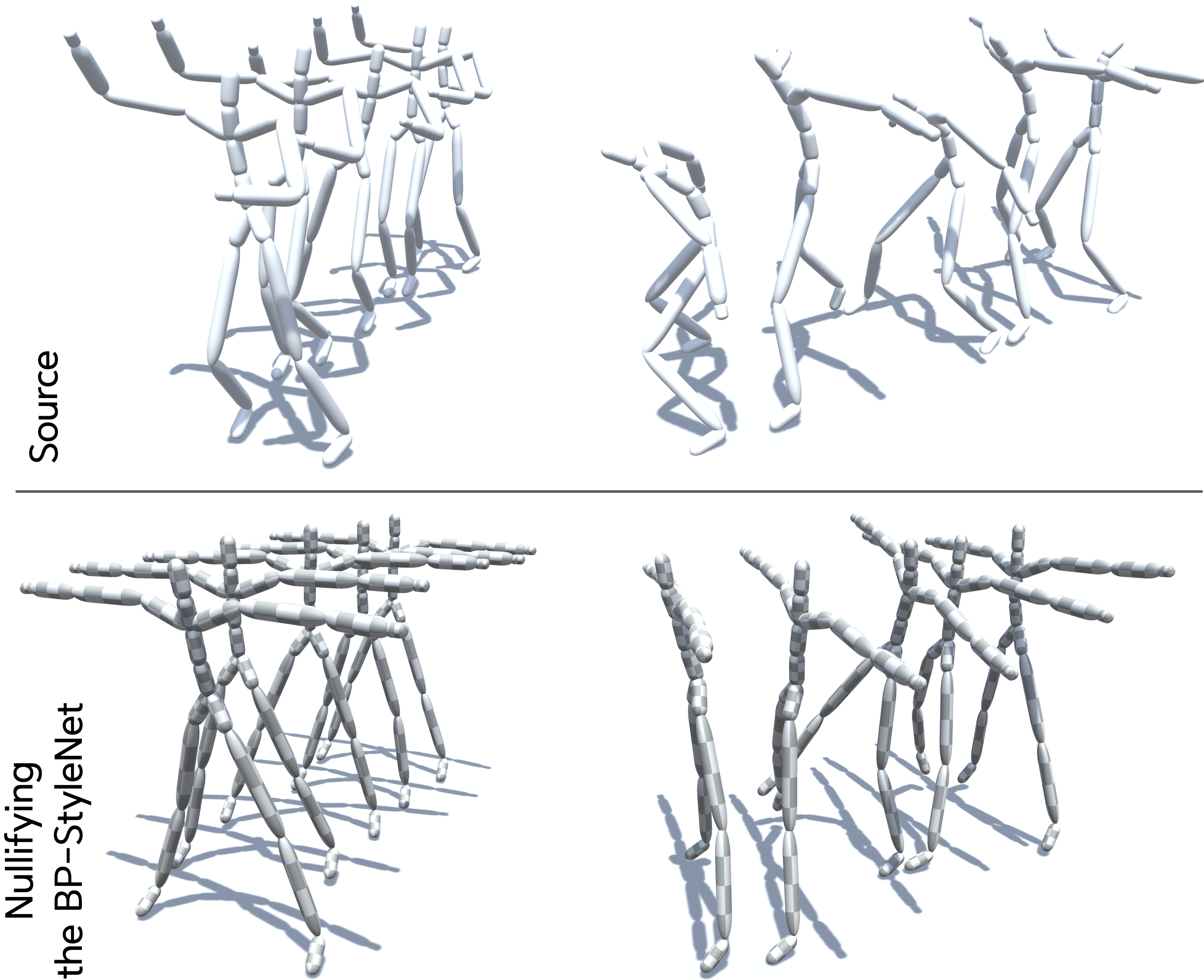}
  \caption{Results of output motions (bottom) only with content features from source motion (top), i.e., without applying style features.}
  \label{fig:content_feature}
\end{figure}

This section examines what information is stored by the content feature and how the attention map in BP-ATN represents the correlation between the style feature and the decoded content feature by visualizing them.

Figure~\ref{fig:content_feature} visualizes the content feature generated from our model. Figure~\ref{fig:content_feature} shows the output motions generated only by the content features obtained from a dancing motion (bottom) and a Pterosauria motion (top). These motions were generated by nullifying the BP-StyleNet by setting the scale and bias of BP-AdaIN to 1 and 0, and the attention map of BP-ATN to 0. 
One can see that the content feature retains the overall walking phase of the source motion while filtering out its style variation of all body parts.
This shows that our content encoder mainly extracts the phase information of each body part from a source motion while other characteristic information of the motion is stored in the style feature.

\begin{figure}[t]
  \centering
  \includegraphics[width=0.8\linewidth]{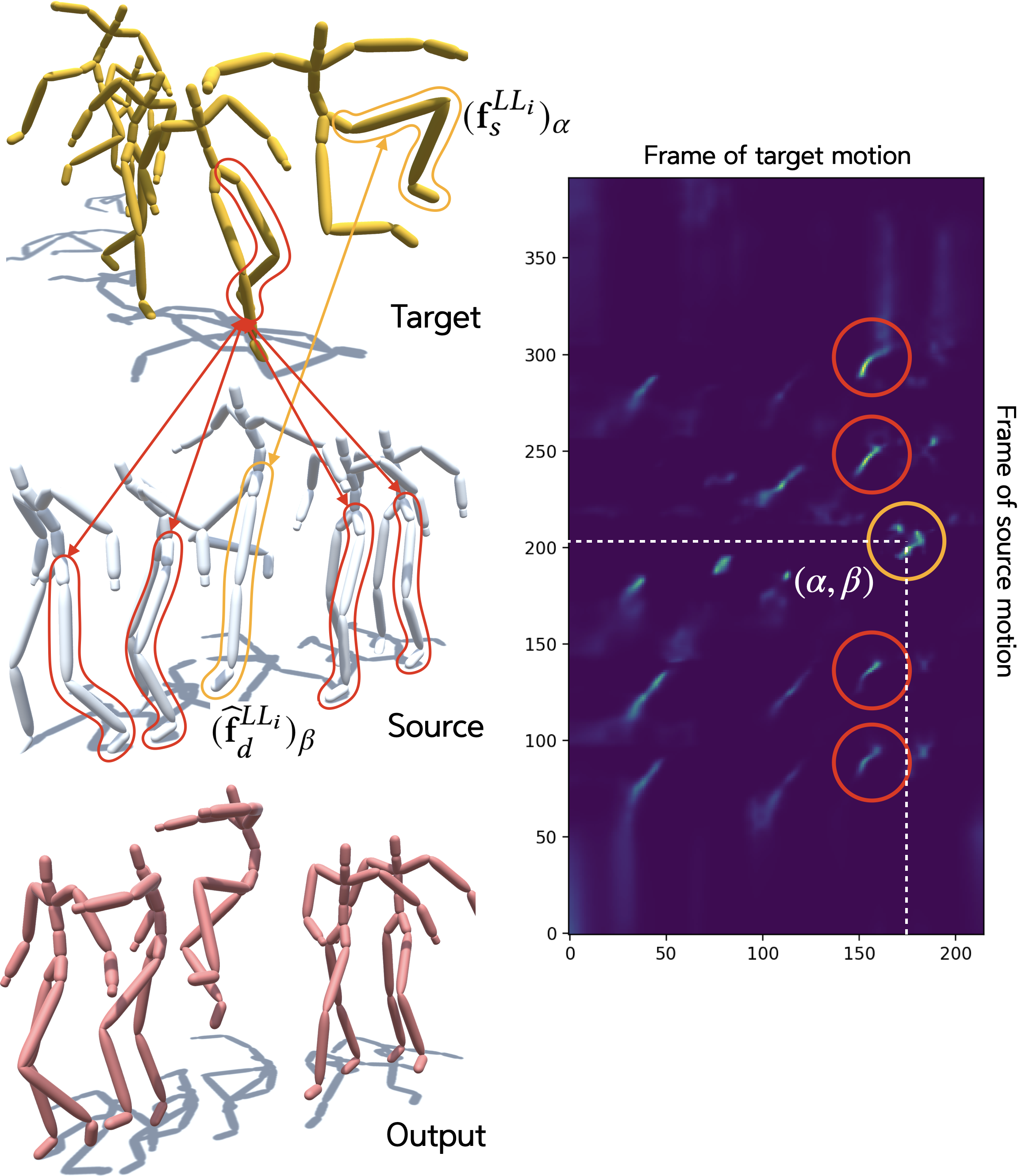}
  \caption{Visualization of the left leg attention map between the decoded content feature from the source motion and the style feature from the target motion (right), and corresponding poses to the marked high attention regions (left).}
  \label{fig:attention_map}
\end{figure}

Next, we visualize attention map $\mathcal{A}$ (Eq.~\ref{eq:attention_map}) that represents the correlation between the style feature $\vf_{s}^{P_i}$  and the decoded content feature $\widehat{\vf}_{d}^{P_i}$. 
Figure~\ref{fig:attention_map} (right) shows the visualization of the attention map of the left leg between a source (jumping while walking) and target motions (repeated swing  jumping). 
To examine the temporal correlation between features in the attention map, the spatial dimension of each feature is averaged out, i.e., $\mathcal{A} \in \real^{T_\text{s} \times T_\text{d}}$ where $X$ and $Y$ axes denote $T_\text{s}$ and $T_\text{d}$.
In the attention map visualization, the higher the attention or correlation, the brighter the color. We identify corresponding poses in the source and target motions to some regions with a high attention value. The yellow circle corresponds to the left leg pose in the middle of jumping in both the source and target motions (yellow borders), which shows that the jumping poses of the source and target motions are found to have a high correlation, promoting the transfer of jumping style of the target to the jumping motion of the source.
In addition, the four red circles correspond to a landing pose in the target motion in a high correlation with four landing poses in the source motion (red borders). This indicates that  periodic poses in a motion can be matched to a single pose in the other motion, and in this case, the landing motion style of the target will be transferred to all landing motions of the source.
%

The output motion shows that the swing jumping style and the landing motion style from the target are well transferred to the jumping motion and the periodic landing motions in the source.
This result suggests that the attention map creates correspondences between similar actions in the source and target motions, driving style transfer between these matched actions. For instance, if a target motion contains walking, jumping, and kicking motions and a source motion contains only a jumping motion. The style transfer would occur from only the jumping motion of the target to the source motion.
By finding the correlation between the content and style features, it significantly contributes to successful motion style transfer, especially capturing time-varying motion style well.

\section{Experiments}
\label{sec:result}
We conduct various experiments to present interesting features and advantages of our method and evaluate its performance.
First, we examine a unique capability of our method, transferring style by body part, with respect to content preservation and per-part stylization. 
Second, we perform a comparison with other previous methods \cite{holden2016deep, aberman2020unpaired} and variations of our model with BP-ATN or BP-StyleNet ablated. 
Lastly, we showcase a real-time motion style transfer combined with an existing character controller.

Motion clips in the test dataset and the new unseen dataset used in the experiment have variable frame lengths, as allowed by our convolutional network-based architecture.
All results are presented after post-processing by the same method in ~\cite{aberman2020unpaired} including foot-sliding removal.
In the figures, the source, target, and resulting motion are shown in white, yellow, and red skeleton, respectively. We used AI4Animation framework~\cite{AI4Animation} for visualizing results.
The supplemental result video shows the resulting motions from the experiments.

\subsection{Motion Style Transfer by Body Part}
\label{subsec:puzzle}

First, we show several examples in Fig.~\ref{fig:teaser} to demonstrate the utility of the per-part style transfer approach. 
Suppose that a zombie walking motion is given, and we want to modify it to make a staggering zombie walking motion. One way would be to add a staggering motion to the trunk and leg while keeping the style of waddling with stretched arm in the source motion. 
Our method realizes this by transferring the style of a staggering motion to only the leg and spine while keeping the style of the given zombie motion for all other body parts as shown in Fig.~\ref{fig:teaser} (the 2nd column). 
In contrast, previous whole-body style transfer methods may replace the zombie style with a staggering style over the whole body. 
Figure \ref{fig:teaser} (the 1st column) shows a case that adds a flapping style to a dance motion by transferring the style of the arm motions of the target to the corresponding parts in the source motion. 
Figure \ref{fig:teaser} (the 3rd and 4th columns) shows the cases of changing the style of parts of a body and the whole body, respectively.

\begin{description}[leftmargin=0pt, itemsep=5.0pt, labelindent=\parindent, listparindent=\parindent] 
\item[Separate transfer to three body parts.]
\begin{figure*}[t]
  \centering
  \includegraphics[width=0.85\textwidth]{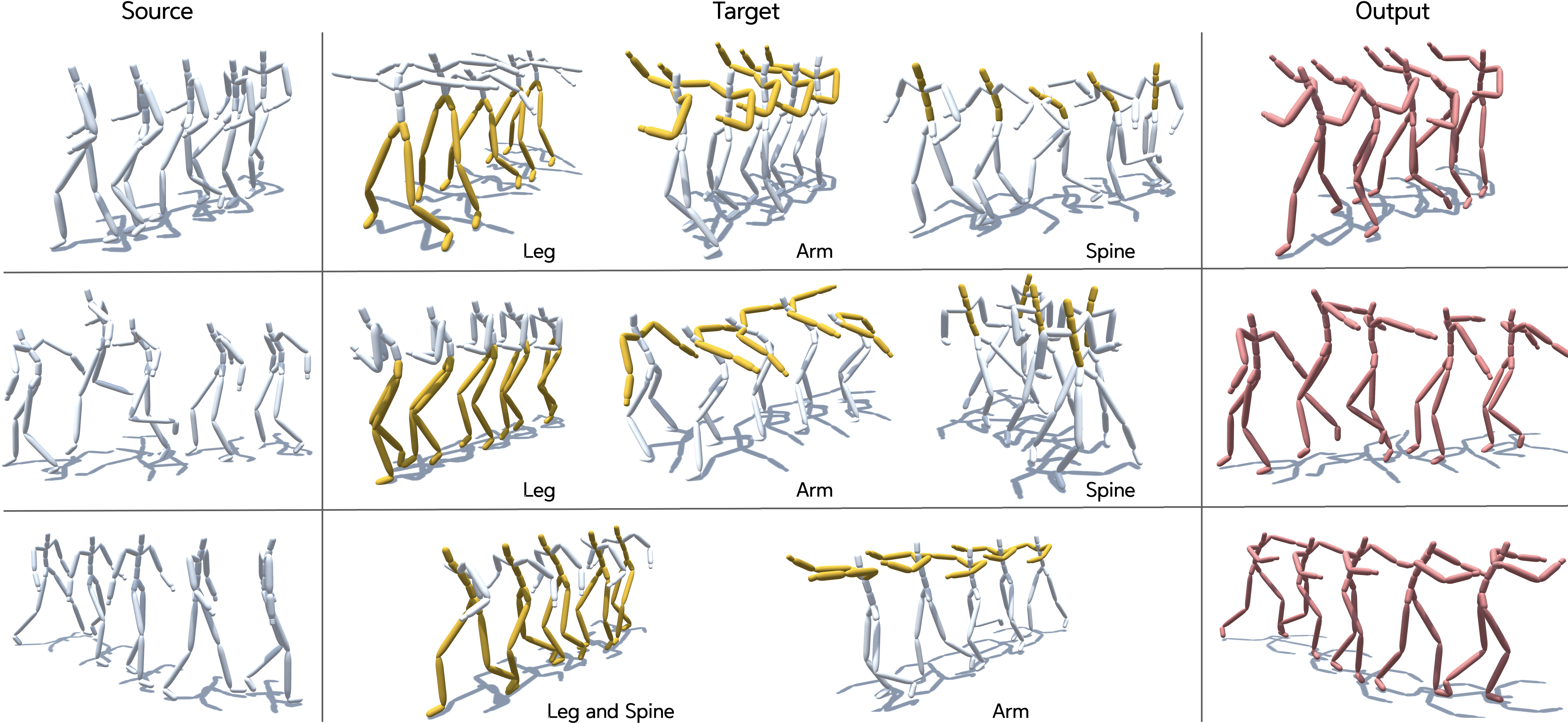}
  \caption{Results of our per-body-part motion style transfer. 
  For each row, two or three separate target motions (middle) for the leg, arm, and spine styles are applied to the source motion (left, gray) to make the output motion (right, red).
  Note that the style code for each target motion is extracted from the target parts marked in yellow.}
  \label{fig:3bodypart}
\end{figure*}

Figure~\ref{fig:3bodypart} shows the results of transferring arbitrary unseen styles from 3 target motions, each for a different body part, to a single source motion.
For convenience, each sub-figure is referred to as $(i,j)$, meaning the j-th image in the i-th row.

First, comparing the source motions and the output motions shows that the contents of the source motions are well preserved in the output motions.
Although all target motions stylizing the leg, arm and spine are clearly different from the source motion, the output motions preserve the phases of walking $(1,5)$, $(3,4)$, and jumping $(2,5)$ motions.
This result suggests that our architecture disentangles content and style well.

In the output motions, each body part shows a realistic motion reflecting the style of its corresponding target motion despite the differences in style and content between target motions. 
For example, the output motion of $(2,5)$ takes the bent leg style of $(2,2)$, the stretched arm style of $(2,3)$ like Pterosaur, and the spine style $(2,4)$ of proud dancing while maintaining the jumping content of the source $(2,1)$ motion. 
Also, the output motion of $(3,4)$ takes the old spine and leg style of $(3,1)$ and the childlike arm style of $(3,3)$ while maintaining the walking content of source $(3,1)$ motion.
This suggests that our per-body-part style transfer is localized well.
The locality is analyzed further next.

\item[Locality.]
\begin{figure}[t]
  \centering
  \includegraphics[width=0.7\linewidth]{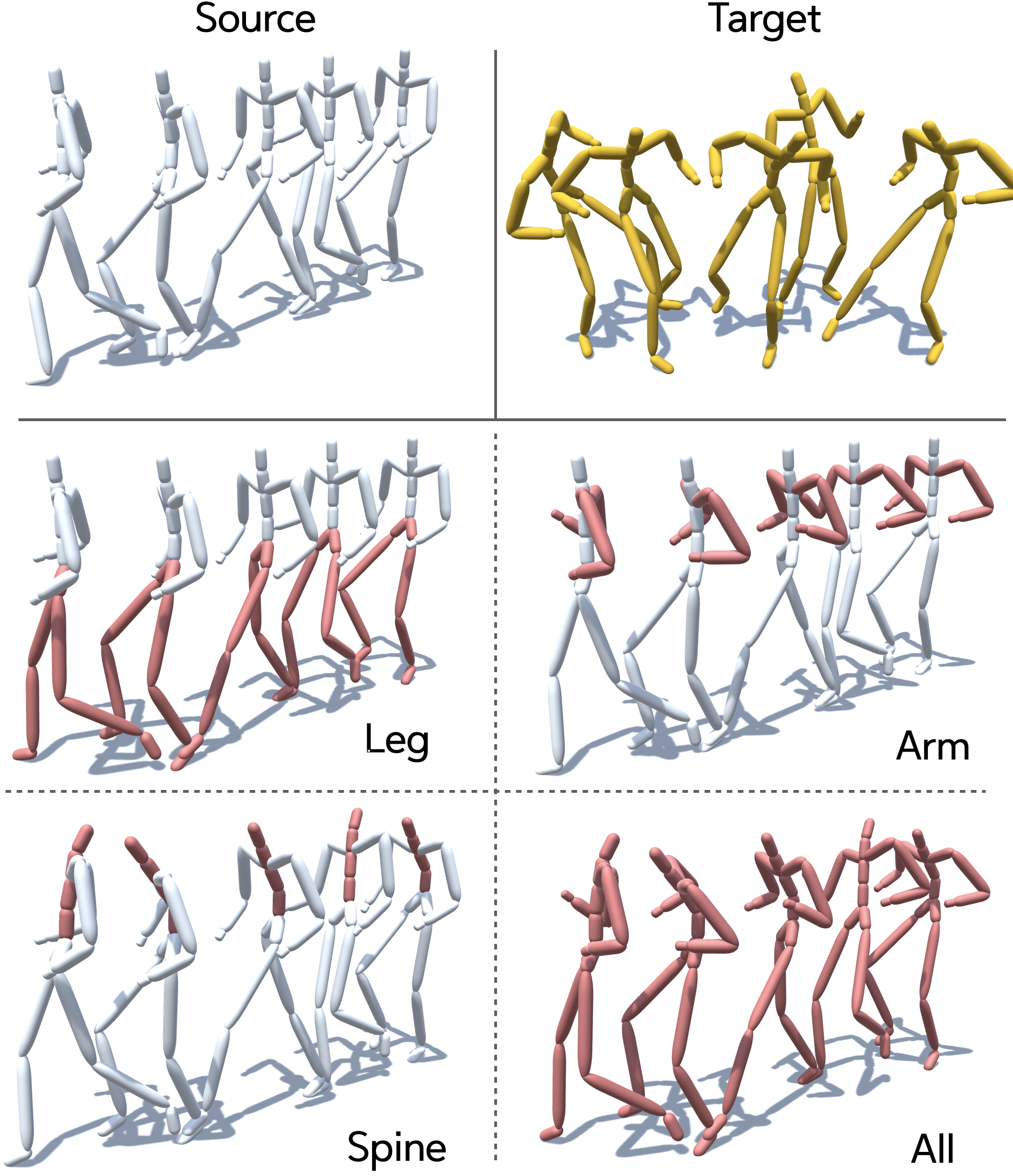}
  \caption{ Locality test. The target style (yellow) is applied only to one body part (red) while keeping the original style with the remaining body parts. The bottom right figure shows the result when the target style is applied to all body parts.}
  \label{fig:locality}
\end{figure}

An essential requirement of our motion style transfer using  \emph{BP-StyleNet} is that its effect is local to each body part while maintaining the content of the source motion.
To validate this, we conduct the following experiment.
We transfer a velociraptor style from a target motion to only one part among the legs, arms and spine of the source motion while keeping the original style for the remaining body parts (Fig.~\ref{fig:locality}). If our style transfer is well localized, when the target style is transferred only to one part, other body parts should keep their original style with only the affected part reflecting the velociraptor style. Figure \ref{fig:locality} shows that our Motion Puzzle framework achieves this.

To quantitatively evaluate the locality, we measure the mean squared displacement (MSD) for each joint between the source motion and four output motions, in which three have only one part stylized (arm, leg and spine) and one has all parts stylized (all) for the experiments in Fig.~\ref{fig:locality}. The more an output motion deviates from the source motion, the larger MSD is. 
Table~\ref{tab:MSD} shows that only the style transferred joints have large MSDs, quantitatively confirming the locality of our method.
The table shows that the shoulder joints (13 and 17) are mildly affected ($0.1 \ge MSD>0.05$) by the spine style transfer, a natural result obtained by the graph convolution across parts.
\begin{table}[t]
    \begin{tabular}{cllll}
    \toprule
    MSD  & \multicolumn{1}{c}{Leg} & \multicolumn{1}{c}{Arm} & \multicolumn{1}{c}{Spine} &    \\ \midrule
    \textgreater{}0.1  & 3, 4, 7, 8   & 15, 16, 19, 20          & 12        \\
    \textgreater{}0.05 & 2, 3, 4, 6, 7, 8    & 14, 15, 16, 18, 19, 20       & 11, 12, 13, 17  \\ \bottomrule
    \end{tabular}
    \caption{The indices of joints with mean squared distance (MSD) of 0.1 or more and 0.05 or more when only one part (leg, arm, or spine) is style transferred for the experiment in Fig.~\ref{fig:locality}. The displacement is measured with respect to the global positions of each joint between source and output motions. Length is normalized by dividing by the skeleton height.
    Joint indices are the leg (1-8), arm (13-20), and spine (0, 9-12).}
    \label{tab:MSD}
\end{table}
%

\item[Per-part style interpolation.]
\begin{figure*}[ht]
  \centering
  \includegraphics[width=0.85\textwidth]{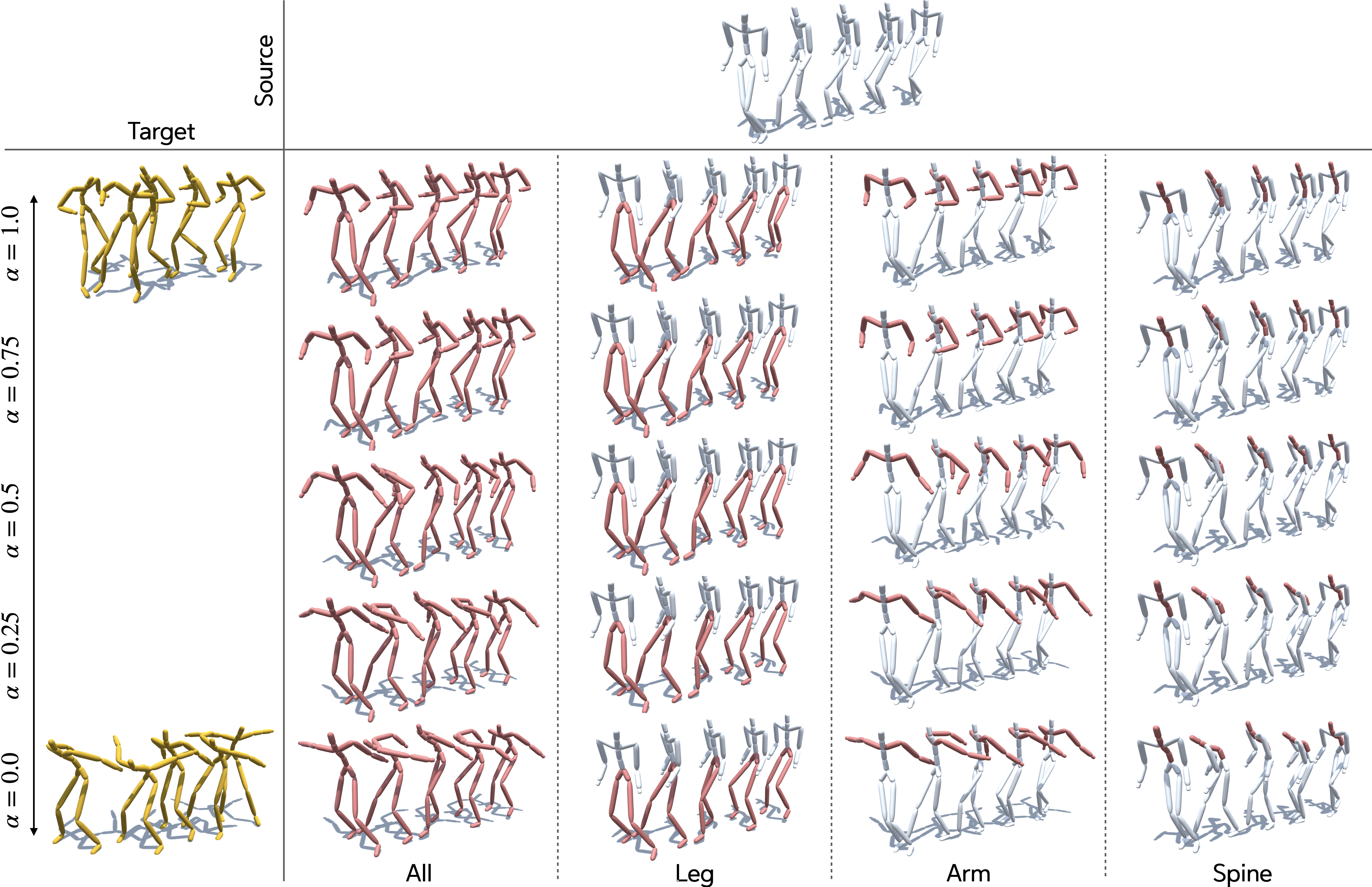}
  \caption{Results of style interpolation by part. Interpolation parameter $\alpha$ of the latent style space varies from 0 (Pteranodon style) to 1 (Velociraptor style).}
  \label{fig:interpolation}
\end{figure*}

This experiment tests whether our style transfer can be interpolated intuitively between two different style motions.
Specifically, two style features $[\vf_s^{G_i}]_{i=1}^3$  and $[{\vf_s^{G_i}}']_{i=1}^3$ from two different target motions are interpolated in the continuous latent style feature space and then applied to the target body part.
Figure~\ref{fig:interpolation} shows the result of transferring interpolated style codes $(1 - \alpha)[\vf_s^{G_i}]_{i=1}^3 + \alpha[{\vf_s^{G_i}}']_{i=1}^3$ from $\alpha=0$ to $\alpha=1$ to each body part.
Since our style latent space is not normalized, linear interpolation was used rather than spherical linear interpolation.
The experiment shows that the interpolation works well while maintaining locality.

\item[Long-term heterogeneous motion.]
\begin{figure}[ht]
  \centering
  \includegraphics[width=0.97\linewidth]{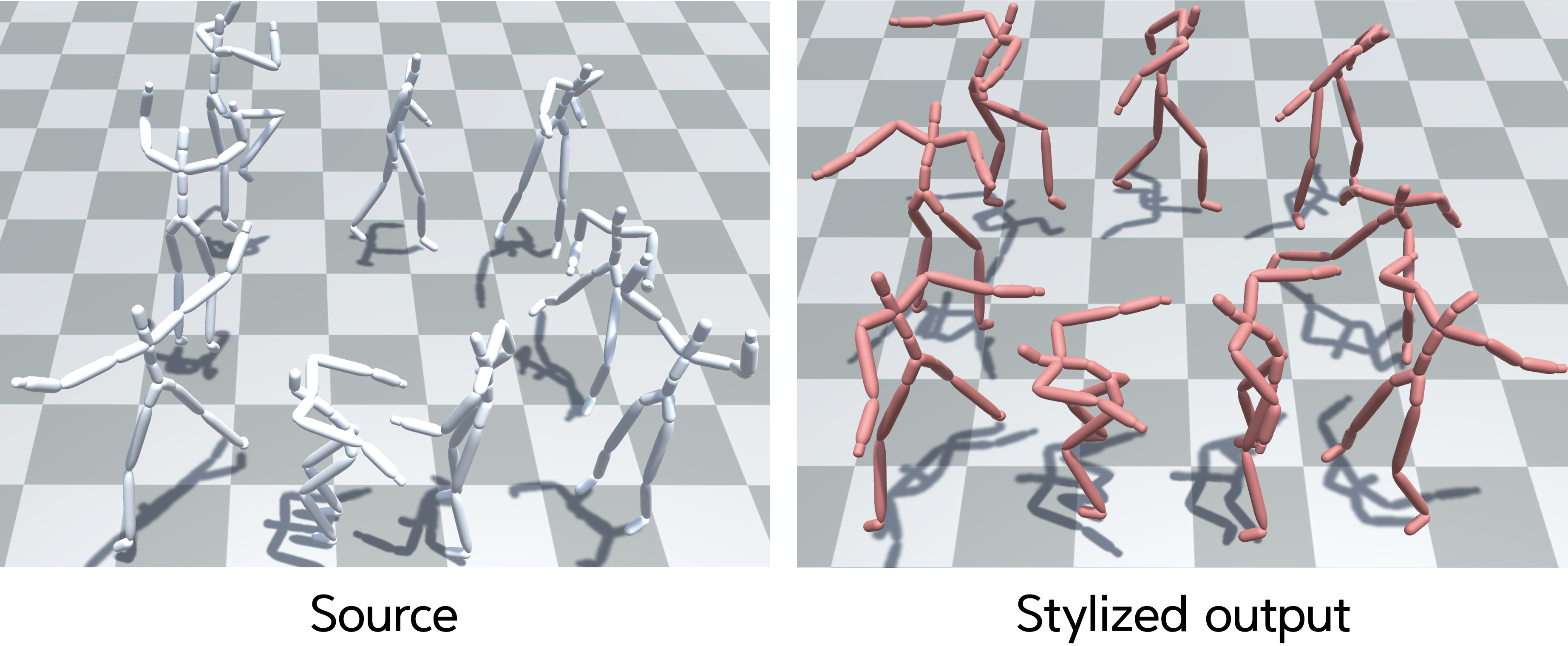}
  \caption{Transferring a Pteranodon style to a long-term heterogeneous source motion containing various behaviors.}
  \label{fig:longterm}
\end{figure}

We test the effectiveness of our framework for long-term heterogeneous motions involving a variety of behaviors. 
Figure~\ref{fig:longterm} shows the whole body style transfer results of a Pteranodon style to a neutral source motion over 3600 frames (1 min), including walking, jumping, transition, kicking and punching. 
We found that our method can be applied robustly to long motions with multiple behaviors.
Additional results of long-term heterogeneous motion with various styles are provided in the supplementary video.

\end{description}

\subsection{Ablation Study and Comparison with Prior Work}
\label{subsec:comparison}

\subsubsection{Qualitative evaluation.}

We conduct an ablation study to examine the effect of BP-StyleNet and BP-ATN modules. In addition, we compare our framework with those in previous work \cite{holden2016deep, aberman2020unpaired}. Since the original framework of \cite{aberman2020unpaired} is effective for style-labeled motion data, we instead construct a comparable framework (dubbed Conv1D+AdaIN) that takes 1D convolution and AdaIN components from \cite{aberman2020unpaired}, and train it using the same losses as ours to enable arbitrary style transfer.

\begin{figure*}[t]
  \centering
  \includegraphics[width=0.8\textwidth]{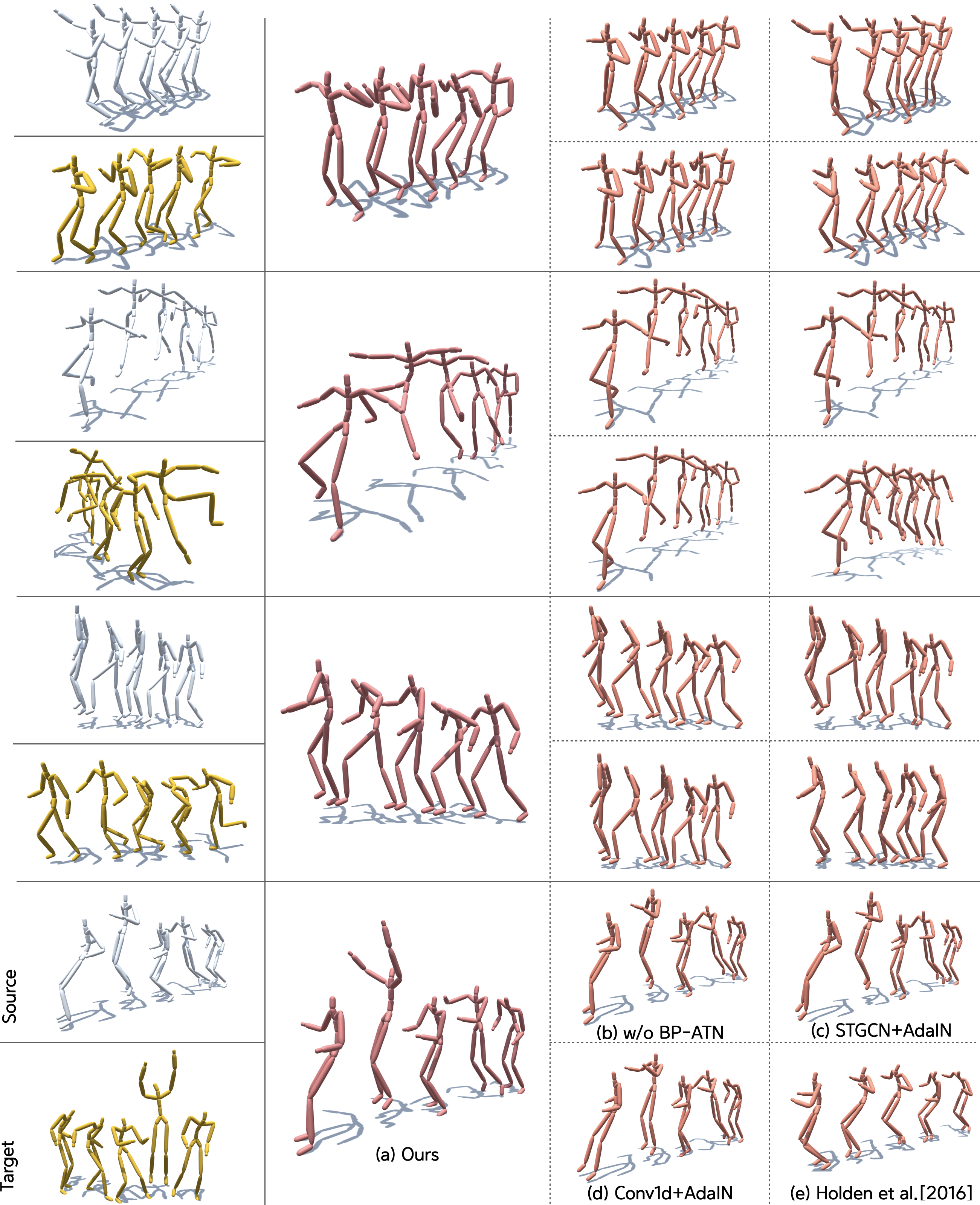}
  \caption{Comparison results of motion style transfer methods. The style of target motions (yellow) are applied to the source motions (grey) to make output motions (our results in red, compared results in orange).
  }
  \label{fig:comparison}
\end{figure*}

Figure~\ref{fig:comparison} shows the comparison of motion style transfer results with our method (a) and its two variations (b and c), Conv1D+AdaIN (d), and a 1D convolution model with Gram matrix as proposed by \cite{holden2016deep} (e). For an ablation study, ``w/o BP-ATN'' (b) is made by removing BP-ATN module from our framework, only using BP-AdaIN to examine the effect of BP-ATN module, and STGCN$+$AdaIN (c) is made by removing BP-StyleNet and performing AdaIN over the whole body, not by body part, to compare BP-AdaIN and AdaIN. Please refer to the supplemental result video to clearly view the differences between the resulting motions.
\paragraph{Effect of BP-ATN}
The results of ``w/o BP-ATN'' (b) in Fig.~\ref{fig:comparison} preserve the content of the source motion well but reflect the style of the target motion less than our results. 
In the second and fourth rows, through the attention map in BP-ATN, jumping parts in the target and source motions are matched to transfer style between corresponding motions. 
The output motions (a) in the first and third rows also show the advantage of BP-ATN.
In the first row, the target motion includes the style of a flapping bent arm, but the output of (b) shows only bent arms that do not flap. Similarly, in the third row, the target motion includes tripping while walking, but the output motion of (b) does not exhibit the tripping style at all and only preserves the content of stair climbing. Since AdaIN simply modifies the mean and variance of the content features to translate the source motion, it captures temporally global features (e.g., bending arms) well but loses temporally local features (e.g., flapping motion). This comparison demonstrates the importance of BP-ATN in transferring local features.
\paragraph{BP-AdaIN vs. AdaIN} 
We found that the style of the target motion is not transferred except for the spine part if AdaIN is used (c) instead of BP-AdaIN for STGCN-based networks. This is because, in the process of transferring style, individual motion styles of each body part are not separated but instead averaged out over the whole body. The overall extent of style transfer is even poorer than Conv1D+AdaIN, a topology-agnostic approach. This result shows that the BP-AdaIN should be used instead of AdaIN for high-quality style transfer for graph-based networks.

As Conv1D+AdaIN (d) uses only AdaIN for style transfer, it reflects the global style features well but may not express temporally local features. In addition, the output motions sometimes jiggle and are distorted, which we attribute to the topology-agnostic 1D convolution scheme. In contrast, the STGCN-based models, (b) and (c), consider human topology, which helps generate plausible motions without motion distortion.

Results of (e) show that the degree of content preservation and the plausibility of motion is lower than the compared models. Since the content and style features are extracted from the same deep features, leading to a dependency between content and style, this approach has a limitation in transferring style between motions with different contents.

\subsubsection{Quantitative evaluation}
\begin{table}[t]
\begin{center}
\caption{Quantitative evaluation on Xia dataset. We calculate FMD, CRA, and SRA on 2500 stylized samples generated by each method on each trial. The table reports the mean ($\pm$ standard deviation) values for each metric over 10 trials.
}
    \begin{tabular}{cccc}
    \toprule
    Methods & FMD$\downarrow$ & CRA$\uparrow$ (\%) & SRA$\uparrow$ (\%)\\
    \toprule
    \text{Real motions} ($\mathcal{M}_\mathrm{gen}$) &  & 96.04 & 90.24 \\
    \midrule
    Ours                        & $\textbf{9.38} \pm 1.10$      & $29.83 \pm 1.35$              & $\textbf{54.94} \pm 2.09$ \\
    Conv1D+AdaIN                & $24.22 \pm 2.29$              & $29.09 \pm 1.29$              & $41.97 \pm 2.01$          \\
    \cite{holden2016deep}       & $29.40 \pm 2.34$              & $\textbf{38.93} \pm 2.09$     & $41.92 \pm 1.77$          \\
    \bottomrule
    \end{tabular}
\label{tab:quantitative}
\end{center}
\end{table}

We quantitatively measure the degree of generation quality, content preservation, and style reflection of three generative models: \cite{holden2016deep}, Conv1D+AdaIN from \cite{aberman2020unpaired}, and ours.
Specifically, we use three metrics; Fr\'echet Motion Distance (FMD), content recognition accuracy (CRA), and style recognition accuracy (SRA). We compute FMD, CRA, and SRA on the stylized set with all possible combinations of source (content) and target (style) motions generated by each model.

The FMD follows the approach of the widely used Fr\'echet Inception Distance (FID) \cite{heusel2017gans}, which is the distance between feature vectors calculated from real and generated images, extracted from Inception v3 network \cite{szegedy2016rethinking} trained on ImageNet~\cite{russakovsky2015imagenet}. It is used to evaluate the quality and diversity of image generative models. 
The FMD measures distance between feature vectors for motion. To this end, we train a content classifier using the method of \cite{yan2018spatial} and use the feature vector obtained from the final pooling layer to measure the FMD between the real and generated motions. A lower FMD suggests higher generation quality.

We use the same content classifier to measure the CRA on generated motions since a model with a higher degree of content preservation would generate more correctly classified motions. The higher CRA indicates better performance with respect to content preservation. 
Similarly, we measure the SRA on stylized sets using another classifier trained to predict the style label of motion.
A higher SRA means better performance for style reflection.

We test generative models using the Xia dataset $\mathcal{M}_\mathrm{test}$~\cite{xia2015realtime}, which is unseen to the models during training. We divide $\mathcal{M}_\mathrm{test}$ into two halves, $\mathcal{M}_\mathrm{cls}$ for classification and $\mathcal{M}_\mathrm{gen}$ for generation. We pre-train the action classifiers with $\mathcal{M}_\mathrm{cls}$ on six content types (\textit{walking, running, jumping, kicking, punching, transitions}) and on eight style types (\textit{angry, childlike, depressed, neutral, old, proud, sexy, strutting}) for content and style classification, respectively. We generate a stylized set using $\mathcal{M}_\mathrm{gen}$ with each generative model. $\mathcal{M}_\mathrm{gen}$ is additionally used as a validation set for the action classifiers and a reference distribution for FMD calculation. 
The pre-trained content and style classifier show 96.04\% and 90.24\% accuracies on $\mathcal{M}_\mathrm{gen}$, respectively, and each stylized set is evaluated by these classifiers.

Table~\ref{tab:quantitative} shows the result. Reasonably, there is a trade-off between the content and style classification. Although the method of \cite{holden2016deep} shows the highest CRA, the style classification is the lowest. 
Conv1D+AdaIN gives a slightly improved SRA with a lower CRA.
Our method increases the SRA performance by a large margin while maintaining a higher CRA than Conv1D+AdaIN. Further, our method outperforms the others in terms of FMD. This result suggests that our method produces higher-quality stylized motions with a greater degree of style reflection while not sacrificing the content preservation performance too much.
The reason why our method shows strong SRA compared with CRA is presumably due to the nature of the content feature. As the content feature mainly includes the phase information while other characteristic aspects are stored in the style feature (Sec.~\ref{sec:discussion}), output motions are strongly affected by the style feature, which may lead to high SRA in comparison with CRA.

\subsubsection{Effect of post-processing}
In the case of using the content feature and style feature from the same motion, reconstructed motion is sufficiently plausible without post-processing. However, combining features from different motions may create foot sliding in the stylized output motion. To remedy this, we compute foot contact labels from the source motion and then use them to correct the output foot positions to maintain the contact phase with inverse kinematics. The target foot position is set as the average foot position in the contact phase. 
The supplementary video shows the comparison between output motions with and without post-processing.


\subsection{Real-time Motion Style Transfer}
\label{subsec:realtime}
\begin{figure}[t]
  \centering
  \includegraphics[width=0.85\linewidth]{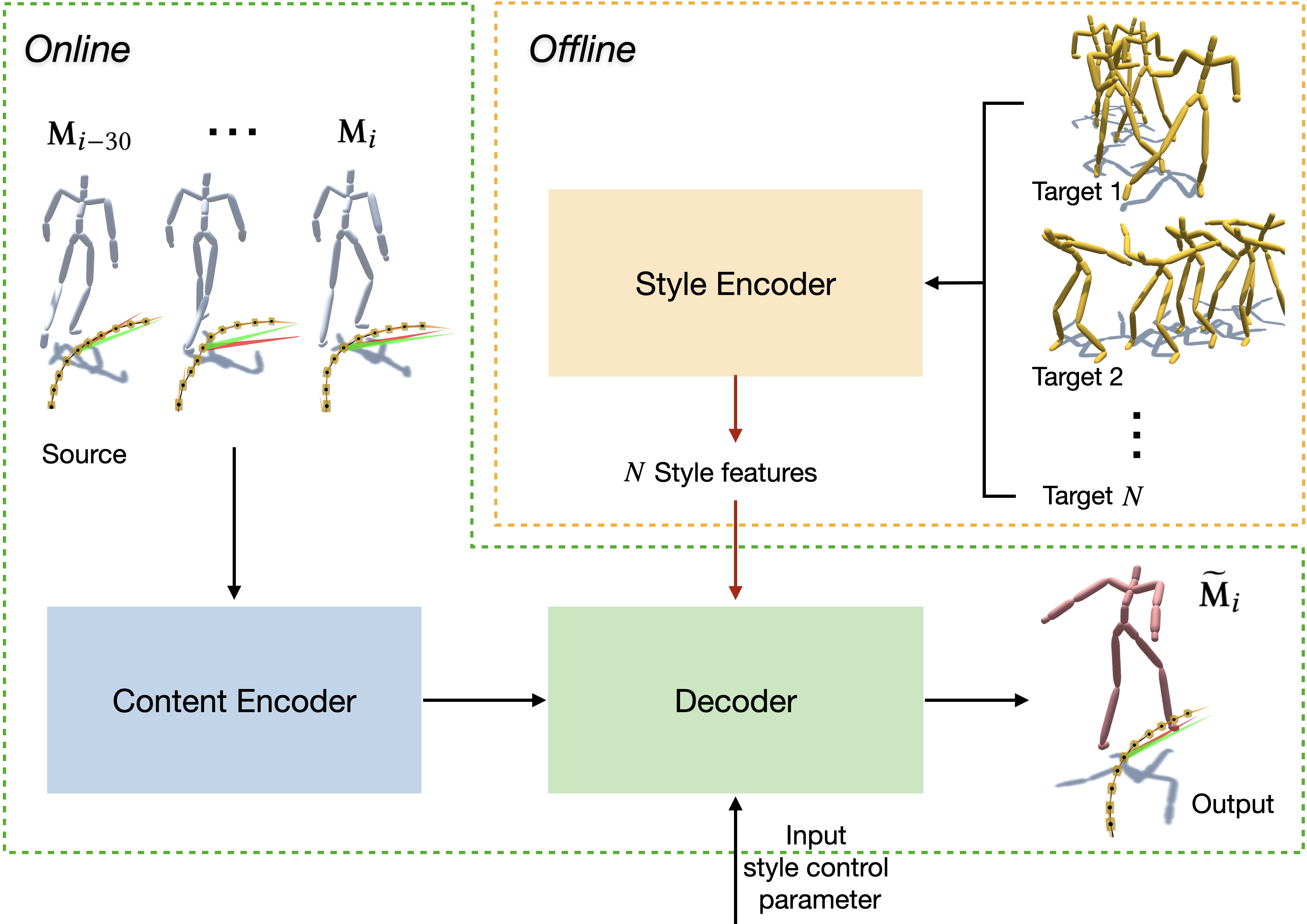}
  \caption{Overview of a real-time motion stylizer.}
  \label{fig:realtime_process}
\end{figure}

\begin{figure}[t]
  \centering
  \includegraphics[width=0.85\linewidth]{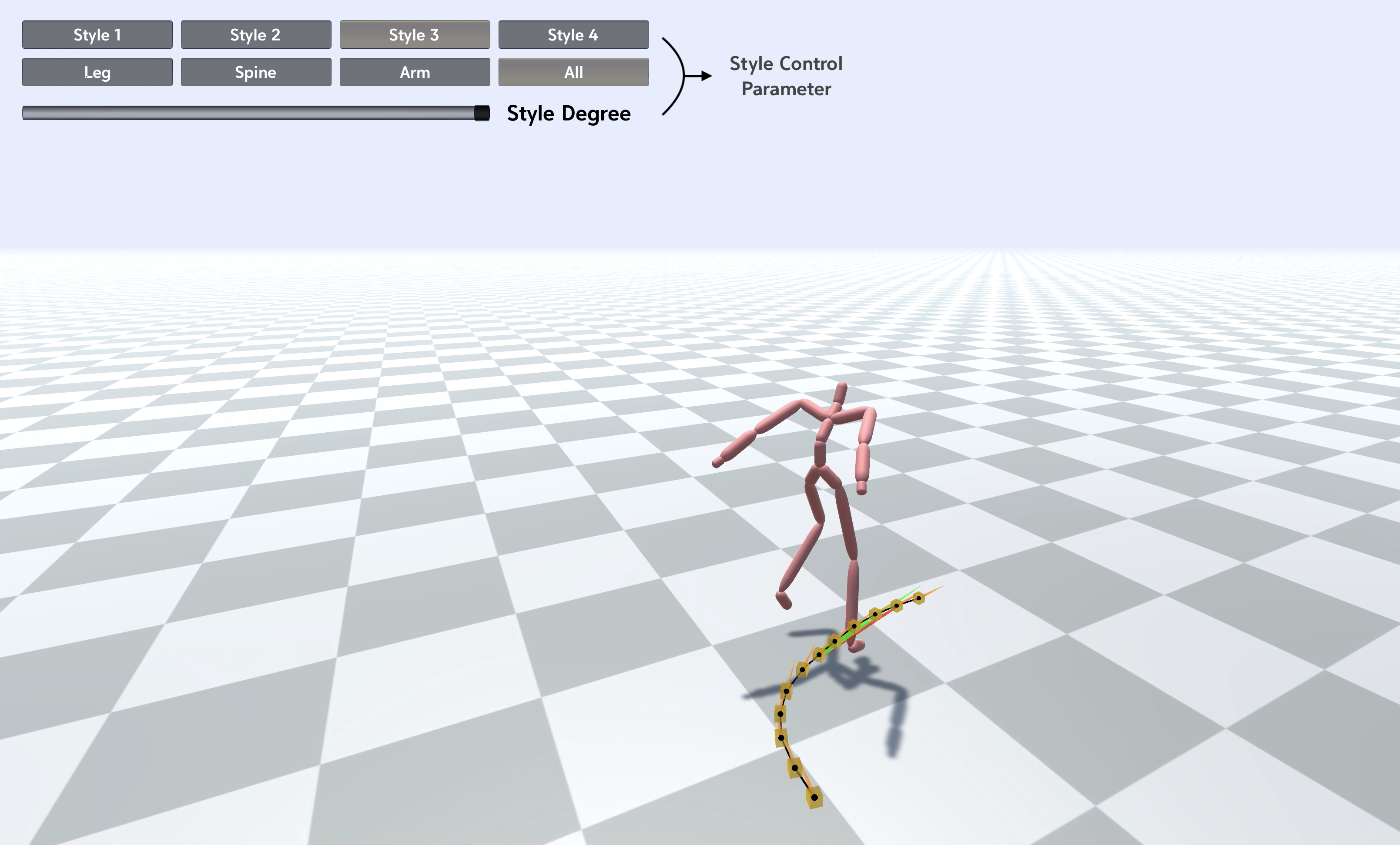}
  \caption{Snapshot of a real-time motion stylizer combined with a PFNN motion controller.}
  \label{fig:realtime}
\end{figure}

In this experiment, we integrate our Motion Puzzle framework with a state-of-the-art motion controller, phase-functioned neural networks (PFNN)~\cite{holden2017phase}, to demonstrate real-time motion style transfer. Our real-time system provides style control parameters to select the target body part, target style, and the degree of style reflection.
To achieve real-time performance, we reduced the required memory size by simplifying the networks while keeping the core structure. For the detailed structure, please refer to Appendix~\ref{Appendix:architecture}.
Figure~\ref{fig:realtime_process} shows the overview of our real-time system to generate the $i$-th frame of the stylized output pose.
In the offline process, our style encoder extracts $N$ (= 4 in our experiment) style features from $N$ target motions and store them.
In the online process, the content encoder extracts content features from the source motion of the duration $[i-30, i]$ (= 0.5sec) generated by PFNN.
When the user selects a target style and target body part, the system applies the selected style feature to the content feature to generate the $i$-th frame pose with the translated style.
Post-processing is not conducted for this experiment as the contact information of the source motion is not available. This real-time motion-style controller runs in 30 FPS. Figure~\ref{fig:realtime} shows a snapshot of the system.

\section{Limitations and Future Work}
\label{sec:limitation}
\begin{figure}[ht]
  \centering
  \includegraphics[width=0.95\linewidth]{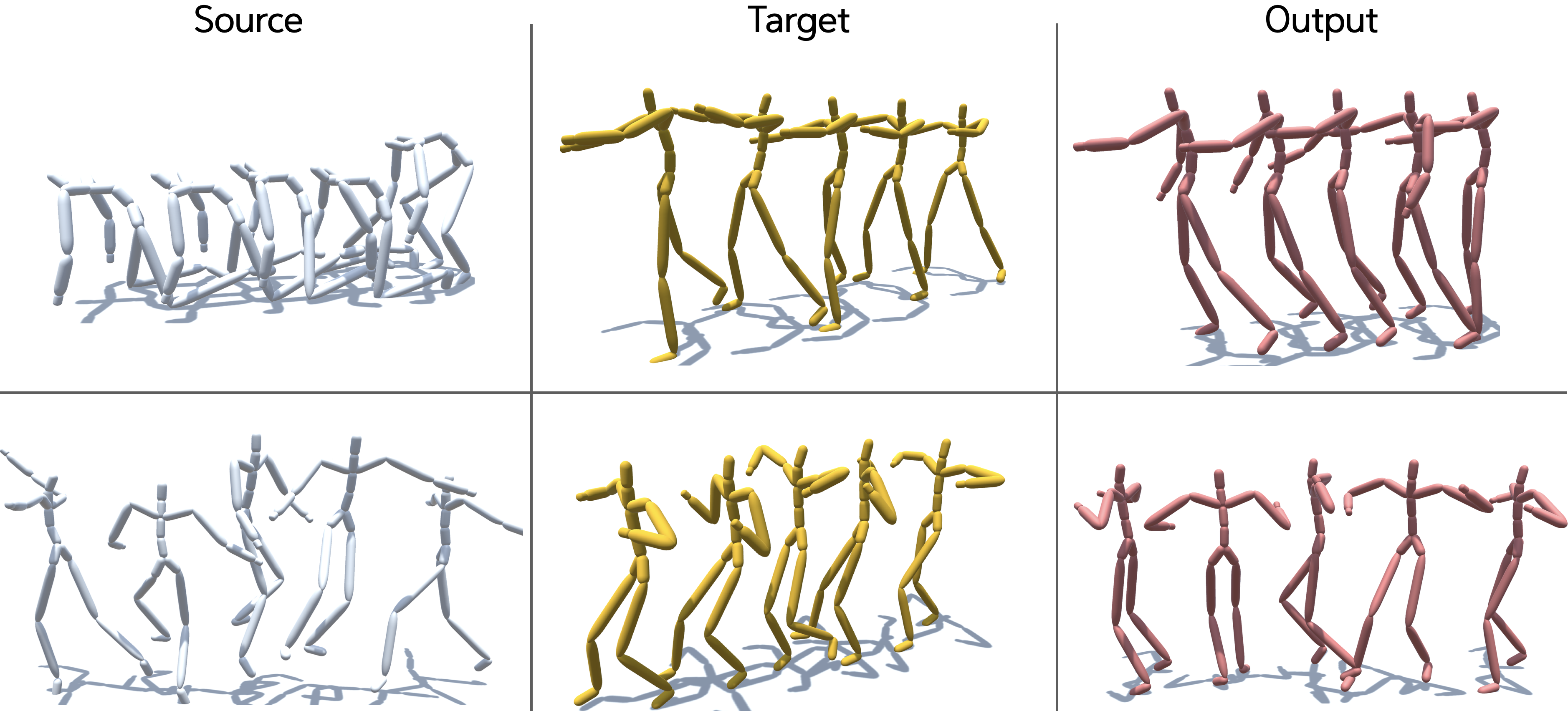}
  \caption{Failure cases. Top: Transferring a zombie style to a crawling source motion breaks the contact between the hand and the ground. Bottom: Rapid rotation in a source motion is collapsed.}
  \label{fig:limitation}
\end{figure}

Our framework has limitations that need to be overcome by future research. 
First, our framework does not consider contact or interaction between the human body and the environment, so the motion content related to contact is not well preserved. For example,  Fig.~\ref{fig:limitation} (top) shows a failure case that the content of crawling in the source motion is not preserved when a zombie walking motion is given as a style target. 
As a future step, we will improve the architecture to include the contact information as an important content feature.

Second, our method is trained to preserve the root motion of the source to the output, which may degrade style transfer quality if the root motion of the target contains important style characteristics, such as dancing. One way to improve this would be to develop an additional module for creating the root velocity to reflect the style of the target motion.

Another limitation is that our framework may not deal with rapidly changing, dynamic motion because it is challenging to identify correspondences between abruptly changing motions within long sequences of the content and target motions. 
Figure~\ref{fig:limitation} (bottom) shows such a case that the output motion fails to preserve the content of rapid rotation in the source motion. A possible solution would be to segment the long motions into sub-sequences and apply the style transfer between the sub-sequences with similar content. 

Since the style of each body part is controlled independently, the resulting whole-body motion may not look coordinated or physically plausible if very different styles are applied over the body parts. An important future work to address this problem would be to add a final step of adjusting the part-stylized motion to improve the naturalness of the motion from holistic viewpoint.

Our framework assumes a single motion as content. We observed that sometimes strong preservation of motion content degrades the effect of stylizing some part, which may be improved if the content component can also be edited. In general, the applicability of a motion editing method will be enhanced if both content and style can be concurrently edited. Thus, an important future research direction would be to control both components by part. A recent technique \cite{Starke2021} that allows per-part motion editing can be a good reference for this direction.

In this paper, we dealt with the motion style of a single person. Solving the problem of stylizing the motion of a group of people, such as collaboration or competition, is another interesting direction for future research. 

\section{Conclusion}
\label{sec:conclusion}
In this paper, we presented the Motion Puzzle framework for motion style transfer. It was carefully designed to respect the kinematic structure of the human skeleton, and thus it can control the motion style of individual body parts while preserving the content of the source motion. 
At the core of the Motion Puzzle framework is our style transfer network, BP-StyleNet, which applies both the global and local traits of target motion style to the source motion.
As a result, our framework can deal with temporally varying motion styles, which was impossible in previous studies.
In addition, our model can be easily integrated with motion generation frameworks, allowing a wide variety of applications, including real-time motion transfer.

\bibliographystyle{ACM-Reference-Format}
\bibliography{reference}

\appendix
\section{Appendix: Implementation}

\subsection{Network architecture}
\label{Appendix:architecture}

\begin{table}[t]
\footnotesize
    \begin{subtable}[h]{0.45\textwidth}
        \centering
        \begin{tabular}{@{}cccccc@{}}
        \toprule
        Layer      & Filter                     & Act.   & Norm.   & Resample & Output shape   \\ 
        \toprule
        $M_\text{tar}$   &              &       &        &         & $T \times 21 \times 15$    \\ 
        \midrule
        Conv1$\times$1 & $\text{k}_s1\text{k}_t1s1$  & -  & -     & -  & $T \times 21 \times 64$    \\
        STConv  & $\text{k}_s3\text{k}_t7s1$ & LReLU  & -       &   & $T \times 21 \times 128$    \\
        STConv  & $\text{k}_s2\text{k}_t5s1$ & LReLU  & -       & Gr-Pool  & $T/2 \times 10 \times 256$ \\
        STConv  & $\text{k}_s2\text{k}_t5s1$ & LReLU  & -       & Gr-Pool  & $T/4 \times 5 \times 512$  \\
        ResBlk  & $\text{k}_s2\text{k}_t3s1$ & LReLU   & -      & -        & $T/4 \times 5 \times 512$  \\ \midrule
        $[\vf_{s}^{G_{i}}]_{i=1}^3$  &         &      &      &         & $[T_i \times J_i \times C_i]_{i=1}^3$    \\ 
        \bottomrule
        \end{tabular}
        \caption{Style Encoder $E_s$.}\label{tab:enc_sty}
    \end{subtable}
    
   \begin{subtable}[h]{0.45\textwidth}
        \centering
        \begin{tabular}{@{}cccccc@{}}
        \toprule
        Layer     & Filter   & Act.  & Norm. & Resample & Output shape   \\ \toprule
        $M_\text{src}$  & -        & -     & -     & -        & $T \times 21 \times 15$    \\ 
        \midrule
        Conv1$\times$1 & $\text{k}_s1\text{k}_t1s1$  & -  & -     & -  & $T \times 21 \times 64$    \\
        STConv  & $\text{k}_s3\text{k}_t7s1$ & LReLU  & IN       &   & $T \times 21 \times 128$    \\
        STConv  & $\text{k}_s2\text{k}_t5s1$ & LReLU  & IN       & Gr-Pool  & $T/2 \times 10 \times 256$ \\
        STConv  & $\text{k}_s2\text{k}_t5s1$ & LReLU  & IN       & Gr-Pool  & $T/4 \times 5 \times 512$  \\
        ResBlk  & $\text{k}_s2\text{k}_t3s1$ & LReLU   & IN      & -        & $T/4 \times 5 \times 512$  \\
        \midrule
        $\vf_{c}^{G_3}$  &         &      &      &         & $T/4 \times 5 \times 512$    \\
        \bottomrule
        \end{tabular}
        \caption{Content Encoder $E_c$.}\label{tab:enc_con}
    \end{subtable}
    
    \begin{subtable}[h]{0.45\textwidth}
        \centering
        \begin{tabular}{@{}cccccc@{}}
        \toprule
        Layer      & Filter        & Resample    & Output shape   \\ 
        \toprule
        $\vf_{c}^{G_3}$      & -    & -     & $T/4 \times 5 \times 512$    \\ 
        \midrule
        BP-ResBlk     & $\text{k}_s2\text{k}_t3s1$    & -          & $T/4 \times 5 \times 512$    \\
        BP-StyleNet   & $\text{k}_s2\text{k}_t5s1$    & -  & $T/2 \times 10 \times 256$ \\
        BP-StyleNet   & $\text{k}_s2\text{k}_t5s1$    & Gr-UnPool     & $T \times 21 \times 128$  \\
        BP-StyleNet   & $\text{k}_s3\text{k}_t7s1$    & Gr-UnPool     & $T \times 21 \times 64$  \\
        Conv$1\times1$          & $\text{k}_s1\text{k}_t1s1$    & -             & $T \times 21 \times 15$  \\
        \midrule
        $\widetilde{M}_\text{src}$  &         &        & $T \times 21 \times 15$    \\
        \bottomrule
        \end{tabular}
        \caption{Decoder $D$.}\label{tab:dec}
    \end{subtable}
    
    \begin{subtable}[h]{0.45\textwidth}
        \centering
        \begin{tabular}{@{}cccccc@{}}
        \toprule
        Layer      & Filter      & Output shape   \\ 
        \toprule
        $\vf_{d}^{G_i}, \vf_{s}^{G_{i}}$   &   & $T_i \times J_i \times C_i$    \\ 
        \midrule
        BP-AdaIN   & -                                    & -                           \\
        LReLU      & -                                    & -                            \\
        STConv     & $\text{k}_s[i]\text{k}_t[i]s1$           & $T_i \times J_i \times C_i$    \\
        BP-ATN     & -                                    & -                            \\
        STConv     & $\text{k}_s[i]\text{k}_t[i]s1$           & $T_i \times J_i \times 2C_i$      \\
        \midrule
        $\widetilde{\vf}_d^{G_i}$   &   & $T_i \times J_i \times C_i/2$  \\
        \bottomrule
        \end{tabular}
        \caption{BP-StyleNet on $G_i$ level.}\label{tab:stylenet}
    \end{subtable}
    
\caption{Architecture of Motion Puzzle framework.}
\label{tab:architecture}
\end{table}

Table~\ref{tab:architecture} shows the details of the network architecture of Motion Puzzle framework. 
The size of spatial-temporal convolutional block (STConv) filter is denoted as $k_s$(\# spatial kernel size)$k_t$(\# temporal kernel size)$s$(\# stride).
The neighbor distances for STConv are $K=(3,2,2)$ for $G_{1},G_{2}$ and $G_{3}$, respectively (Eq.~\ref{eq:sampling}).
All graph pooling (Gr-Pool) and graph unpooling (Gr-UnPool) are performed before each layers.

The style encoder $E_s$ consist of 3 STConv with 2 Gr-Pool, followed by a STConv residual block (ResBlk).
$E_s$ extracts style features for each level graph $G_i$ from target motion $M_\text{tar}$ and produces $[\vf_{s}^{G_{i}}]_{i=1}^3$. The output dimension of style features $\vf_{s}^{G_{i}}$ are $T_1 \times J_1 \times C_1 = T \times 21 \times 128$, $T_2 \times J_2 \times C_2 = T/2 \times 10 \times 256$ and $T_3 \times J_3 \times C_3 = T/4 \times 5 \times 512$.
Here, residual block follows the form used in \cite{liu2019few}. 

The content encoder $E_c$ receives a source motion as input, passes it through 3 STConv and 2 Gr-Pool layers, followed by a STConv residual block (ResBlK) to generate content feature $\vf_c^{G_3}$. Every STConv and ResBlk layer has Instance Normalization (IN) for removing style variation. Note that we only extract content feature for $G_3$ level as mentioned in Sec.~\ref{subsec:con_en}.

The decoder $D$ has roughly a reverse structure of the encoder. 
For each level, BP-StyleNet applies a corresponding style feature to the decoded content feature. Details of BP-StyleNet are provided in Table~\ref{tab:architecture}(c). 

The BP-StyleNet on $G_i$ level consists of BP-AdaIN, BP-ATN and 2 STConv between them. 
The BP-StyleNet receives a decoded content feature $\vf_d^{G_i}$ and a style feature $\vf_s^{G_i}$ to generate a translated feature $\widetilde{\vf}_d^{G_i}$.
Output shape  $T_i \times J_i \times C_i$ and $\text{k}_s[i]\text{k}_t[i]s1$ are noted in each level of decoder part.

For the real-time motion style transfer in Sec. \ref{subsec:realtime}, we simplified the networks to reduce the memory size.
We removed the residual blocks from the encoders and decoder and applied BP-StyleNet only in $G_3$ and $G_2$ levels. And i-frame translated pose is obtained by slicing in [i-30, i] translated motion.
As a result, we reduced the memory size by more than 80$\%$ while maintaining a suitable quality in the translated output motion. 

\subsection{Training Details}
\label{Appendix:training}
Our architecture is trained for 10 epochs with a batch size of 32. Learning rates for all networks are set to $10^{-4}$.
Training time is about 5 hours with two NVIDIA GTX 2080ti.
All networks are implemented with PyTorch.
We set $\lambda_\text{cyc}=1$, $\lambda_\text{root}=1$ and $\lambda_\text{sm}=1$ in Eq.~\eqref{eq:loss_total}.
We use the RAdam optimizer~\cite{liu2019variance} with $\beta_1=0$ and $\beta_2=0.99$.
For better results, we adopt the exponential moving averages~\cite{karras2018progressive} over all parameters of every network.

\end{document}